\documentclass[a4paper,prb,aps,showpacs,twocolumn,superscriptaddress]{revtex4-1}
\usepackage	[T1]			{fontenc}		
\usepackage	[latin1]		{inputenc}		
\usepackage				{graphicx, epstopdf, epsfig, rotating, floatpag}
\usepackage	[caption=false]	{subfig}
\usepackage				{booktabs, tabu, longtable, dcolumn}
\usepackage				{amsmath, amssymb, amsfonts, nicefrac, xfrac, gensymb, braket, bm} 
\usepackage				{paralist, enumerate, lipsum}
\usepackage				{color, verbatim}
\usepackage				[english]	{babel}
\usepackage				{varioref}
\usepackage				{units}
\usepackage	[pdftex, bookmarks, colorlinks, breaklinks]
						{hyperref}																
\hypersetup				{colorlinks=true, linkcolor=black, citecolor=blue, filecolor=dullmagenta, urlcolor=blue}	

\graphicspath	{	
			}

\begin{document}

\title{Single-ion anisotropy and magnetic field response in spin ice materials Ho$_{2}$Ti$_{2}$O$_{7}$ and Dy$_{2}$Ti$_{2}$O$_{7}$}

\author	{Bruno Tomasello}
\affiliation	{SEPnet and Hubbard Theory Consortium, University of Kent, Canterbury CT2 7NH, U.K.}
\affiliation	{ISIS facility, STFC Rutherford Appleton Laboratory, Harwell Oxford Campus, Didcot OX11 0QX,  U.K.}

\author	{Claudio Castelnovo}
\affiliation	{TCM group, Cavendish Laboratory, University of Cambridge, Cambridge CB3 0HE, U.K.}

\author	{Roderich Moessner}
\affiliation	{Max-Planck-Institut f\"{u}r Physik komplexer Systeme, 01187 Dresden, Germany}

\author	{Jorge Quintanilla}
\affiliation	{SEPnet and Hubbard Theory Consortium, University of Kent, Canterbury CT2 7NH, U.K.}
\affiliation	{ISIS facility, STFC Rutherford Appleton Laboratory, Harwell Oxford Campus, Didcot OX11 0QX,  U.K.}

\begin{abstract}

Motivated by its role as a central pillar of current theories of dynamics of spin ice in and out of equilibrium, we study the single-ion dynamics of the magnetic rare earth ions in their local environments, subject to the effective fields set up by the magnetic moments they interact with. This effective field has a transverse component with respect to the local easy-axis of the crystal electric field, which can induce quantum tunnelling. We go beyond the projective spin-1/2 picture and use instead the full crystal-field Hamiltonian. 
We find that the Kramers vs non-Kramers nature, as well as the symmetries of the crystal-field Hamiltonian, result in different perturbative behaviour at small fields ($\lesssim 1$~T), with transverse field effects being more pronounced in Ho$_{2}$Ti$_{2}$O$_{7}$ than in Dy$_{2}$Ti$_{2}$O$_{7}$.
Remarkably, the energy splitting range we find is consistent with time scales extracted from experiments. 
We also present a study of the static magnetic response which highlights the anisotropy of the system in the form of an off-diagonal $g$ tensor and we investigate the effects of thermal fluctuations in the temperature regime of relevance to experiments. 
We show that there is a narrow yet accessible window of experimental parameters where the anisotropic response can be observed. 

\end{abstract}

\maketitle

\section{	Introduction	\label{sec:Intro} 	}

The properties and behaviour of spin ice materials, Ho$_{2}$Ti$_{2}$O$_{7}$ (HTO) and Dy$_{2}$Ti$_{2}$O$_{7}$ (DTO) among others, are deeply rooted in the characteristic single-ion anisotropy of rare earth (RE) magnetism. 
The two lowest energy states (degenerate at single-ion level) are separated by a large energy gap ($\gtrsim 200$~K) from the other excited states, thus projecting the system onto an effective spin-$1/2$ space at low temperatures. The lowest energy doublet has moreover a strong easy-axis anisotropy, which is responsible for its classical Ising-like behaviour~\cite{Bramwell:2001} (for a recent detailed discussion, see also Ref.~\onlinecite{Rau:2015}).
These properties justify modelling the magnetic moments as classical Ising spins with a local easy axis. 
The rich thermodynamic behaviour of spin ice systems can be largely accounted for by the physics of the ground state doublet combined with the pyrochlore lattice structure and exchange and dipolar interactions: frustration leads to an extensively degenerate ground state~\cite{Bramwell:2001}, topological order, and an emergent gauge symmetry hosting magnetic monopole excitations~\cite{Castelnovo:2008,Castelnovo:2012}. 

This thermodynamic model of spin ice was later promoted to a dynamical one by introducing an experimentally inspired~\cite{Snyder:2001,Ehlers:2003,Snyder:2004,Matsuhira:2011} single spin-flip time scale~\cite{Jaubert:2009} (see also Ref.~\onlinecite{Ryzhkin:2005}). This choice was motivated by the experimental observation of a well-defined microscopic time scale in the magnetic response of these materials, which appears to be largely temperature independent in the regime of interest. 
Such dynamical modelling of spin ice proved reasonably successful at capturing the experimental response and relaxation properties, and triggered a new research direction into the behaviour of these systems out of equilibrium -- an interesting and highly tuneable setting that combines topological properties, kinematic constraints, emergent point-like quasiparticles and long-range Coulombic interactions~\cite{Castelnovo:2010,Mostame:2014}. 

A temperature-independent microscopic spin-flip time scale is typically associated with quantum tunnelling under an energy barrier that the ion has to traverse in order to reverse its magnetic polarisation. 
Understanding this behaviour clearly requires that we go beyond the single-ion ground-state doublet (spin-$1/2$) approximation, and we investigate the role of possible quantum perturbations that may be responsible for the tunnelling dynamics. 
To date, such understanding appears to be lacking in the literature. 

The work presented here is a step at gaining insight into the quantum single-ion dynamics in spin ice HTO and DTO. Specifically, we focus on spin-spin interactions as a source of quantum fluctuations. The exchange and dipolar fields acting on a given ion due to others in the system have both a longitudinal and a transverse component with respect to the local easy axis. The latter acts as a \emph{transverse field} in the effective Ising model.
We study in detail the effects of such transverse field on the single-ion behaviour, obtaining the resulting energy splitting (namely, inverse characteristic time scales) and anisotropic response, both at zero as well as finite temperature. 

There exists a concrete motivation for studying the specific case of an exclusively transverse magnetic field, 
a setting which at first sight seems to require fine-tuning the longitudinal component to vanish. 
This may not appear straightforward in spin ice, a dense assembly of large rare earth moments interacting via a long-range and geometrically complex dipolar interactions, Eq.~\eqref{eq:HamIsingDipolar}. 

However, spin ice is no stranger to such fine-tuning. It is now well understood how the geometry of the pyrochlore lattice conspires with that of
the dipolar interaction to ensure that the longitudinal total field on each spin is, to a good approximation, 
equal for {\it all spins} in {\it all ground states}~\cite{Isakov:2005}, 
which are exponentially numerous~\cite{Ramirez:1999} and in general not related by any symmetry transformation.

Similarly, a pointlike defect in a spin ice ground state, known as magnetic monopole~\cite{Castelnovo:2008} 
has an energy independent of its location, as long as it is spatially well separated from other monopoles. 
As the spatial displacement of a monopole proceeds via the flip of a single spin, this spin must be subject to a vanishing longitudinal field 
-- otherwise the excitation energy in the system, encoded in that of the monopole, would change as the monopole moves. 

Therefore, our study can be thought of as providing a picture of the quantum mechanics underpinning the motion of an isolated monopole defect in a ground state of spin ice. 
The properties of such mobile monopoles are currently subject to both experimental~\cite{Paulsen:2014, Pan:arX2015, Tokiwa:arX2015} and theoretical work~\cite{Petrova:arX2015}.

For the purpose of the present paper, we approximate the exchange interactions by their classical form. 
Namely, we consider the interaction Hamiltonian 
\begin{equation}
\begin{split}
	H	=&	\, - J \sum_{\langle ij\rangle}	
			\vec{S}_{i} \cdot \vec{S}_{j}								\\[0.1cm]
		&
			+	Dr_{nn}^{3}	\sum_{(ij)}		\Bigg[
												\frac	{\vec{S}_{i} \cdot \vec{S}_{j}}
												{|\mathbf{r}_{ij}|^{3}}
												-	\frac	{3(\vec{S}_{i}\cdot\mathbf{r}_{ij})(\vec{S}_{j}\cdot\mathbf{r}_{ij})}
												{|\mathbf{r}_{ij}|^{5}}
										\Bigg]
		\, ,
\end{split}
\label{eq:HamIsingDipolar}
\end{equation}
where $i,j$ label the sites of the pyrochlore lattice; $\vec{S}_{i} = \sigma_i \,\mathbf{z}_{i}$, with $\sigma_{i}=\pm 1$ and $\mathbf{z}_{i}$ are the (4 inequivalent) unit vectors pointing from one tetrahedral sublattice to the other; 
$J$ and $D$ are the exchange and dipolar coupling constants, respectively; 
$r_{nn}$ is the nearest neighbour distance on the pyrochlore lattice; and 
$r_{ij} $ is the distance between the two sites $ i $ and $ j $. 
Within the approximation of this Hamiltonian, the action of all other ions on a given one is an effective magnetic field, 
\begin{equation}
\begin{split}
	H	=&	\, \sum_{i}	 \vec{B}_{\rm eff}(i) \cdot \vec{S}_{i}
\nonumber 
\\[0.1cm]
	\vec{B}_{\rm eff}(i)	=&	\, - J \sum_{j,\: \langle ij\rangle}	
			\vec{S}_{j}								\\[0.1cm]
		&
			+	Dr_{nn}^{3}	\sum_{j}		\Bigg[
											\frac	{\vec{S}_{j}}
											{|\mathbf{r}_{ij}|^{3}}
											-	\frac	{3(\vec{S}_{j}\cdot\mathbf{r}_{ij})\, \mathbf{r}_{ij}}
											{|\mathbf{r}_{ij}|^{5}}
										\Bigg]
		\, ,
\end{split}
\label{eq:HamIsingDipolar eff field}
\end{equation}
whose strength and direction were studied in Ref.~\onlinecite{Sala:2012}. Here we thus limit ourselves to considering the action of an applied field on the full single-ion Hamiltonian, beyond the customary projection to its lowest lying states. 

We find that the Kramers vs non-Kramers nature of HTO vs DTO results in a different perturbative behaviour at small fields, whereby in HTO the ground state doublet splits at second order in the applied field strength, whereas DTO only splits at third order, as illustrated in Fig.~\ref{fig:Splitting}. One therefore expects transverse fields in HTO to be more effective at inducing quantum tunnelling dynamics than in DTO (at small fields, $\lesssim 0.5$~T). 
Using degenerate perturbation theory, we provide an analytical understanding of this difference in behaviour in terms of symmetries of the CF Hamiltonian.  
Remarkably, the energy splitting range we find is consistent with quantum tunnelling time scales observed in experiments~\cite{Snyder:2004}. 

We also present a detailed study of the static magnetic response to a transverse field, which highlights the anisotropy of the system. We find interesting resonances as a function of the in-plane direction of the field, where off-diagonal components of the $g$-tensor become non-vanishing (namely, a purely transverse field in the $xy$-plane induces a longitudinal response along the local $z$ axis). 

We investigate the effects of thermal fluctuations in the temperature regime of relevance to experiments. We find that in thermal equilibrium much of the anisotropic response averages out up to rather large fields ($\sim$ a few Tesla) for temperatures as low as a hundred milliKelvin. 
Nonetheless, signatures of the anisotropic response in spin ice could be experimentally observed at low temperatures in fields $\sim 10$~T. 
Our results further support the robustness of the classical easy-axis Ising approximation for the single-ion behaviour in spin ice, while at the same time helping to quantify its limit of validity. 

We stress that all of the above features can only be grasped via a full description of the single-ion Hamiltonian capturing the complexity of its interaction with the other spins / environment. They cannot be understood (but at most added at an effective level) if we limit our modelling projectively to the lowest CF levels. 

The paper is organised as follows. 
Sec~\ref{sec:CFsingleion} introduces the full single-ion crystal-field (CF) Hamiltonian 
for rare earth ions Ho$^{3+}$ and Dy$^{3+}$ in spin ice. 
Sec.~\ref{sec:MagnField} investigates the effects of a magnetic field at zero temperature, with specific focus on fields transverse to the local easy axis. 
We use exact diagonalisation (Sec.~\ref{sec:MagnField_ExactDiag}) as well as 
degenerate perturbation theory in the limit of small fields 
(Sec.~\ref{sec:MagnField_PerturbTheory}). 
Thanks to the large CF energy scales typical of these systems, perturbation theory is indeed valid well into the range of field strengths of experimental interest. 
Finally, Sec.~\ref{sec:FiniteTemp} discusses thermal effects in the relevant temperature range and studies the behaviour of the resulting single-ion magnetic susceptibility, 
and in Sec.~\ref{sec:Discussion} we summarise and discuss our results. 
 
\section	{	crystal-field of spin ice RE$^{3+}$ ions			\label{sec:CFsingleion}		}

The general formula for spin ice pyrochlore oxides is 
A$^{3+}_{2}$B$^{4+}_{2}$O$^{2-}_{7}$, 
where the A and B species are rare earth (RE) and transition metal (TM) cations, respectively~\cite{Subramanian:1993, Gardner:2010, AbragamBleaneyBook:1987}.
The structure is given by the space group $F_{d\bar{3}m}$  
featuring two sub-lattices that interpenetrate each other and consist of networks of corner-sharing tetrahedra. 
In HTO and DTO the A magnetic sites host, respectively, the Ho$^{3+}$ and Dy$^{3+}$ ions, 
while the B sites are occupied by non-magnetic Ti$^{4+}$ ions.

\begin{figure}[ht!]
	\centering
	\subfloat	[Perspective view		\label{fig:CFenvironment(a)}]	
					{\includegraphics[trim=	0mm		0mm		0mm		0mm, clip, width=.9\linewidth]{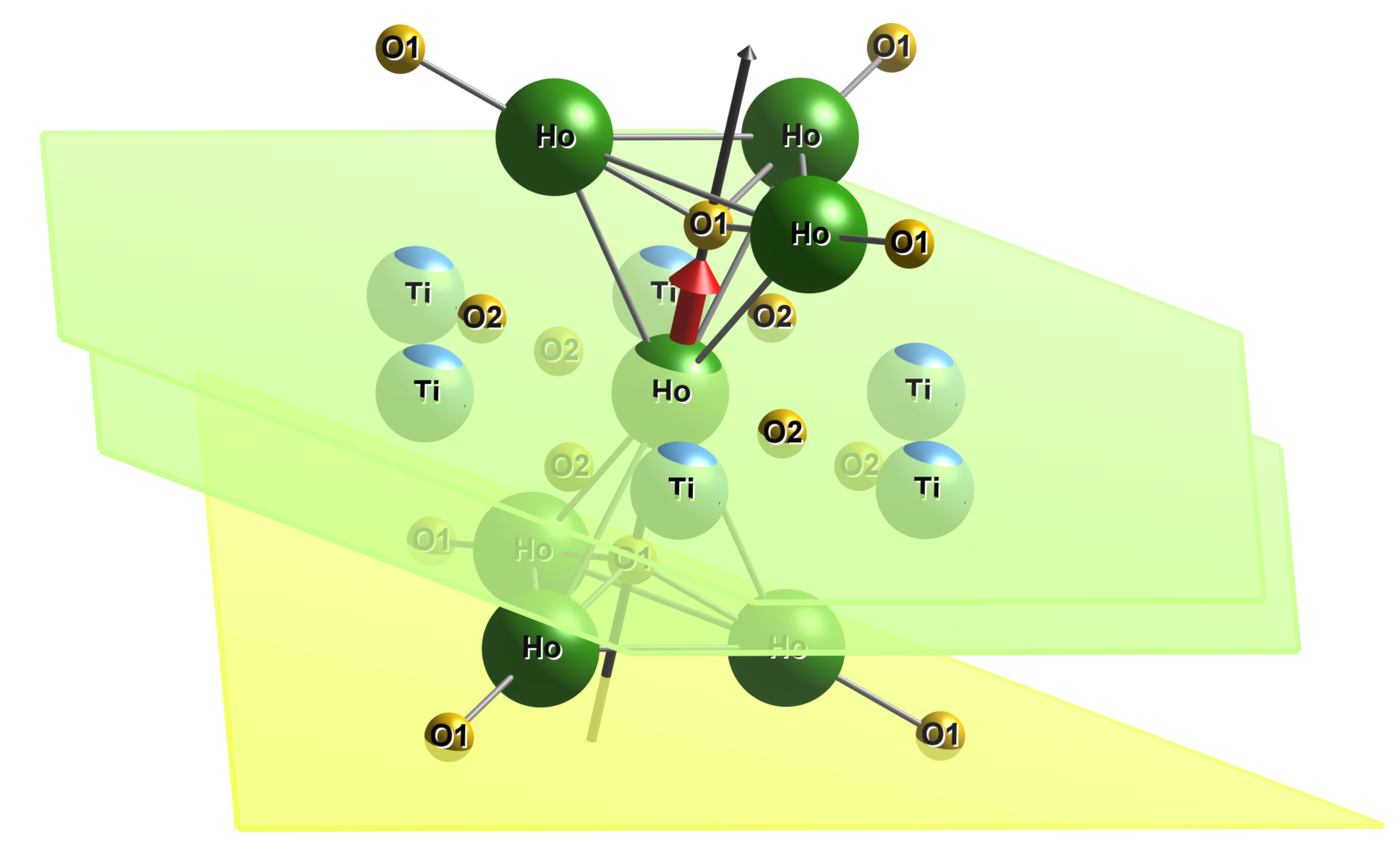}}

	\subfloat	[Side view					\label{fig:CFenvironment(b)}]
					{\includegraphics[trim = 0mm	-60mm		0mm 	0mm,	clip, width=.43\linewidth]{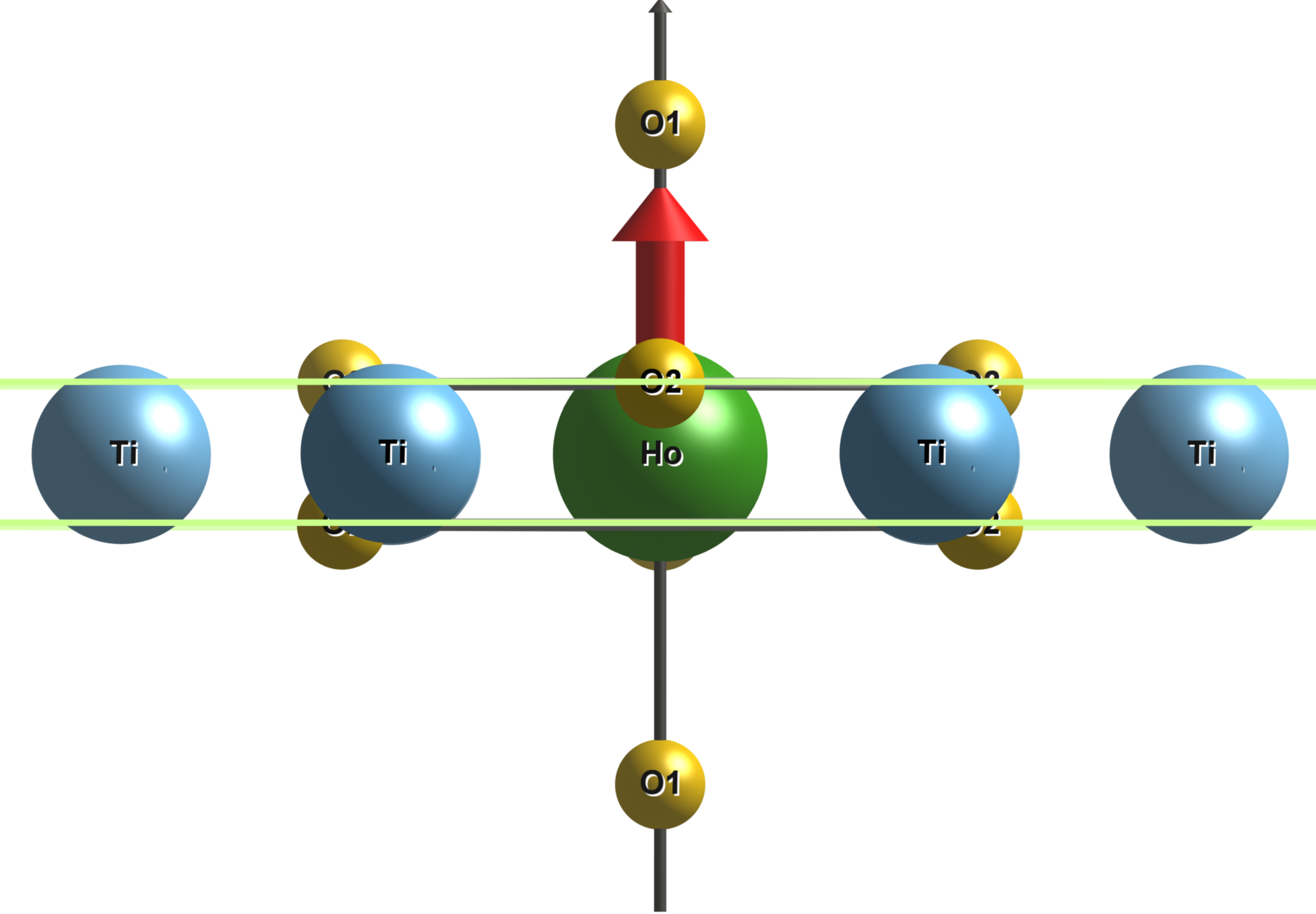}}
	\quad
	\subfloat	[Top view					\label{fig:CFenvironment(c)}]
				{\includegraphics[trim=	0mm		0mm		0mm		0mm, clip, width=.47\linewidth]{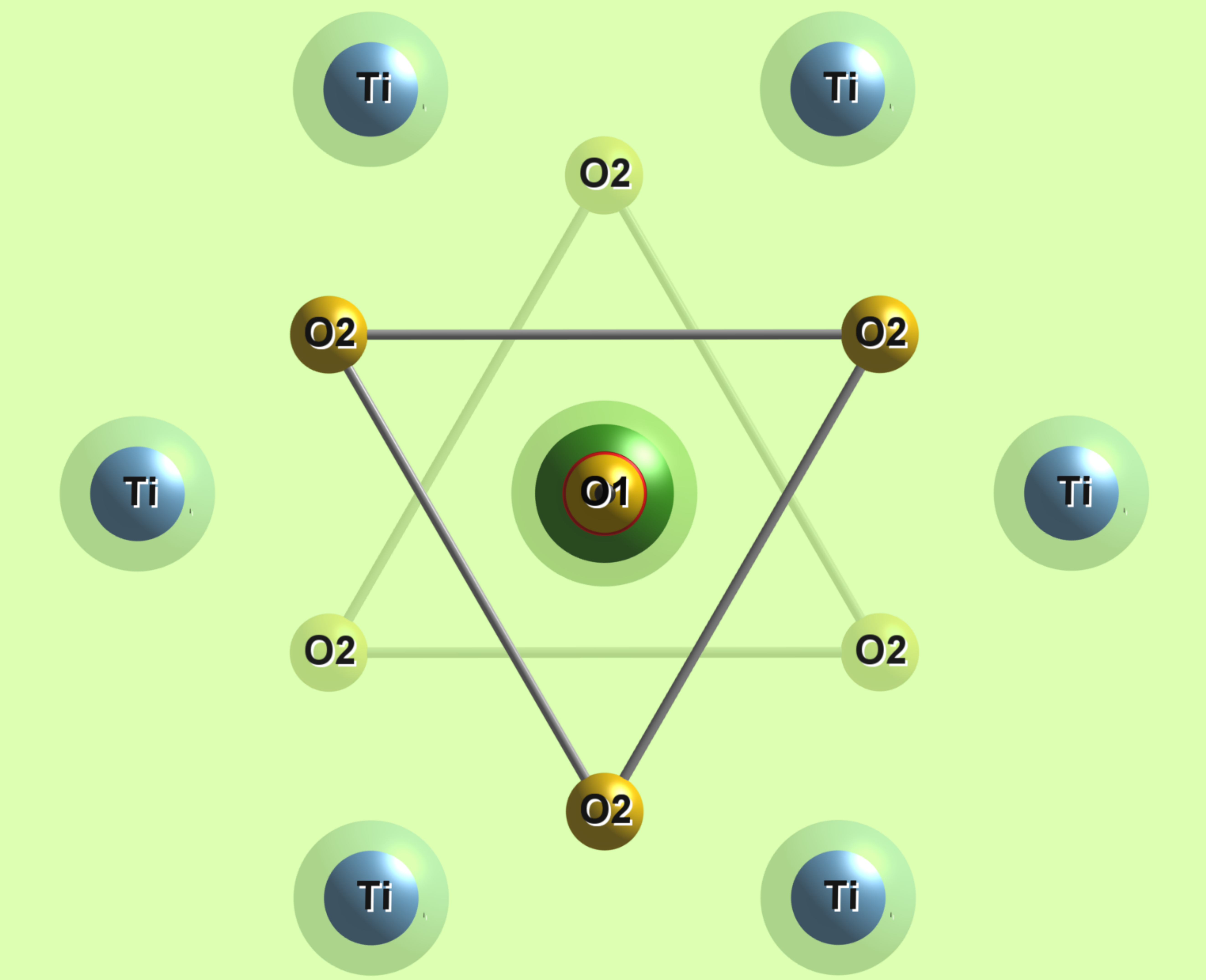}}
	\caption	{	\label{fig:CFenvironment}
				The crystal-field environment of a RE$^{3+}$ ion in a magnetic pyrochlore oxide. Ho$^{3+}$ is used for concreteness.
				Panels (a), (b) and (c) show respectively a tilted, side and top view of the same structure. For clarity, panel (a) additionally displays the six surrounfing Ho$^{3+}$ ions and their second axial oxygens. 
                                The Ti$^{4+}$ ions are arranged in an hexagon coplanar with the RE$^{3+}$ ion in the centre.
				The edges of the triangles connect the coplanar O2 oxygens: three above and three below 
				the plane of the RE$^{3+}$ and Ti$^{4+}$ ions.
				The two green planes shown are parallel to each other and contain the respective triangles of O2 ions.
				The antiprismatic arrangement of the six O2 gives the $D_{3d}$ point-group symmetry.
				The two remaining O1 ions, aligned along the $\langle111\rangle$ axis with the central RE$^{3+}$, drive the local Ising anisotropy.
			}
\end{figure}

The local point group symmetry for the RE$^{3+}$ ions in magnetic pyrochlore oxides is a trigonal $D_{3d}$ (see App.~\ref{sec:App_MagnPyro}). 
This is schematically shown in Fig.~\ref{fig:CFenvironment} and it accounts for the arrangement of the eight oxygen ions (yellow spheres) surrounding the rare earth ion (green sphere).
The oxygen sites are distinguished in two main subclasses according to their position with respect to the central RE-site: the O1 sites and the O2 sites.
The strong axial alignment of the O1 ions (above and below the central RE$^{3+}$ ion) drives the classical Ising-like anisotropy typical of spin ice materials.
The O2 ions, displaced in equilateral triangles lying in parallel planes transverse to the easy axis of the O1 ions, are responsible for the antiprismatic character of the $D_{3d}$ symmetry. 

The crystal-field Hamiltonian of a rare earth ion in a $D_{3d}$ symmetry can be conveniently expressed as~\cite{Hufner:1978,Takegahara:2000} 
\begin{equation}
	\begin{split}
	\hat{\mathcal{H}}_{\mathrm{CF}} =	&	\widetilde{B}_0^2 \, \hat{O}_0^2		+	\widetilde{B}_0^4 \, \hat{O}_0^4 		+	\widetilde{B}_3^4 \, \hat{O}_3^4		\\[0.1cm]
															+	&	\widetilde{B}_0^6 \, \hat{O}_0^6		+	\widetilde{B}_3^6 \, \hat{O}_3^6		
															+		\widetilde{B}_6^6 \, \hat{O}_6^6 ,
	\end{split}
	\label{eq:HamCFpyroStevensShort}
\end{equation}
where the Stevens operators $\hat{O}_{q}^{k}$
together with the respective parameters $ \widetilde{B}_{q}^{k} $ determine the CF spectrum and eigenfunctions of each compound. 
Following the general convention~\cite{Hufner:1978}, the $ \hat{O}_k^q $ are such that 
the $ k=0 $ operators are $ q $-polynomials of only diagonal operators $ \hat{\mathbf{J}}^2, \hat{J}_{z}$,
while those with $ k>0 $ include also $ k $-powers of the ladder operators $ \hat{J}_{+},\hat{J}_{-} $. 
A list of the matrix elements of the Stevens operators in the $ \ket{J,M_{J}}$ basis, where $ J,M_{J} $ are the quantum numbers for the total angular momentum and its projection along the local $\langle111\rangle$ axis respectively, is given in Ref.~\onlinecite{Hutchings:1964}, and can be straightforwardly obtained from their operator expressions in App.~\ref{sec:App_MagnPyro_SOps}. 
The crystal-field parameters for HTO and DTO are listed in  Table~\ref{tab:spiniceCFparamStev}. 

\begin{table}[ht]	\centering
\setlength{\tabcolsep}{0.3cm}
\def\arraystretch{1.5}%
\begin{tabular}{ c | c c |}
									&			HTO	(meV)	&			DTO	(meV)		\\[0.1cm]	\hline
			$\widetilde{B}_0^2$	&		$-7.6	\times 10^{-2}$	&		$-1.6	\times 10^{-1}$		\\[0.1cm]	\hline
			$\widetilde{B}_0^4$	&		$-1.1	\times 10^{-3}$	&		$-2.3	\times 10^{-3}$		\\[0.1cm]	\hline	
			$\widetilde{B}_3^4$	&		$8.2	\times 10^{-3}$	&		$1.6	\times 10^{-2}$		\\[0.1cm]	\hline	
			$\widetilde{B}_0^6$	&		$-7.0	\times 10^{-6}$	&		$6.5	\times 10^{-6}$		\\[0.1cm]	\hline
			$\widetilde{B}_3^6$	&		$-1.0	\times 10^{-4}$	&		$9.9	\times 10^{-5}$		\\[0.1cm]	\hline
			$\widetilde{B}_6^6$	&		$-1.3	\times 10^{-4}$	&		$1.0	\times 10^{-4}$		\\
			\hline
\end{tabular}
	\caption	{	
				The crystal-field parameters (in meV) for the Hamiltonian in Eq.~\eqref{eq:HamCFpyroStevensShort} obtained from Refs.~\onlinecite{Rosenkranz:2000, Malkin:2010} (see also App.~\ref{sec:App_MagnPyro_CF}).  
			}
\label{tab:spiniceCFparamStev}						
\end{table}						

\begin{figure}
	\centering
	\subfloat	[\label{fig:CFenergiesHTO}]
					{\includegraphics[trim=	0mm		0mm		0mm		0mm, clip, width=.48\linewidth]{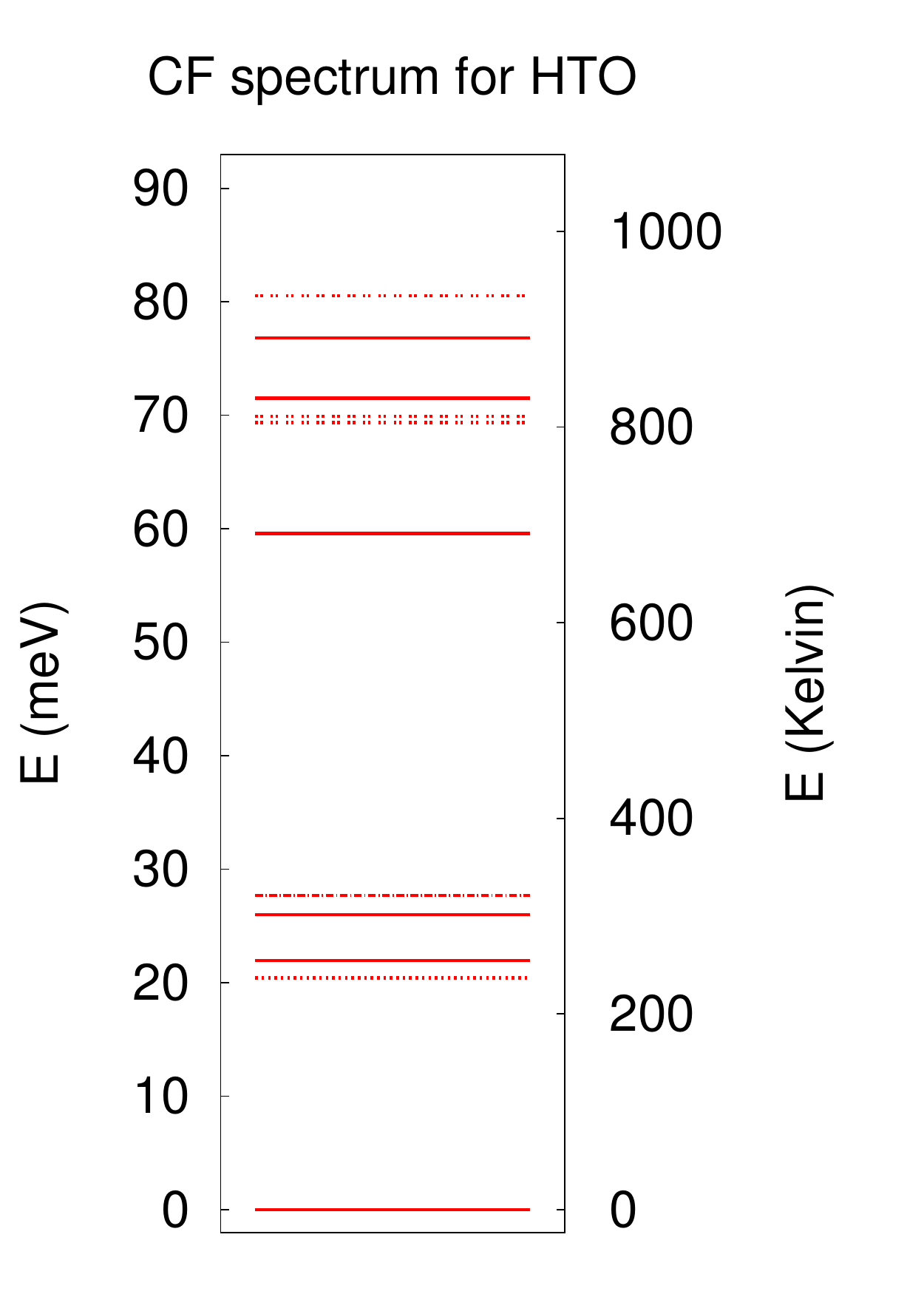}}
	\quad
	\subfloat	[\label{fig:CFenergiesDTO}]
					{\includegraphics[trim=	0mm		0mm		0mm		0mm, clip, width=.48\linewidth]{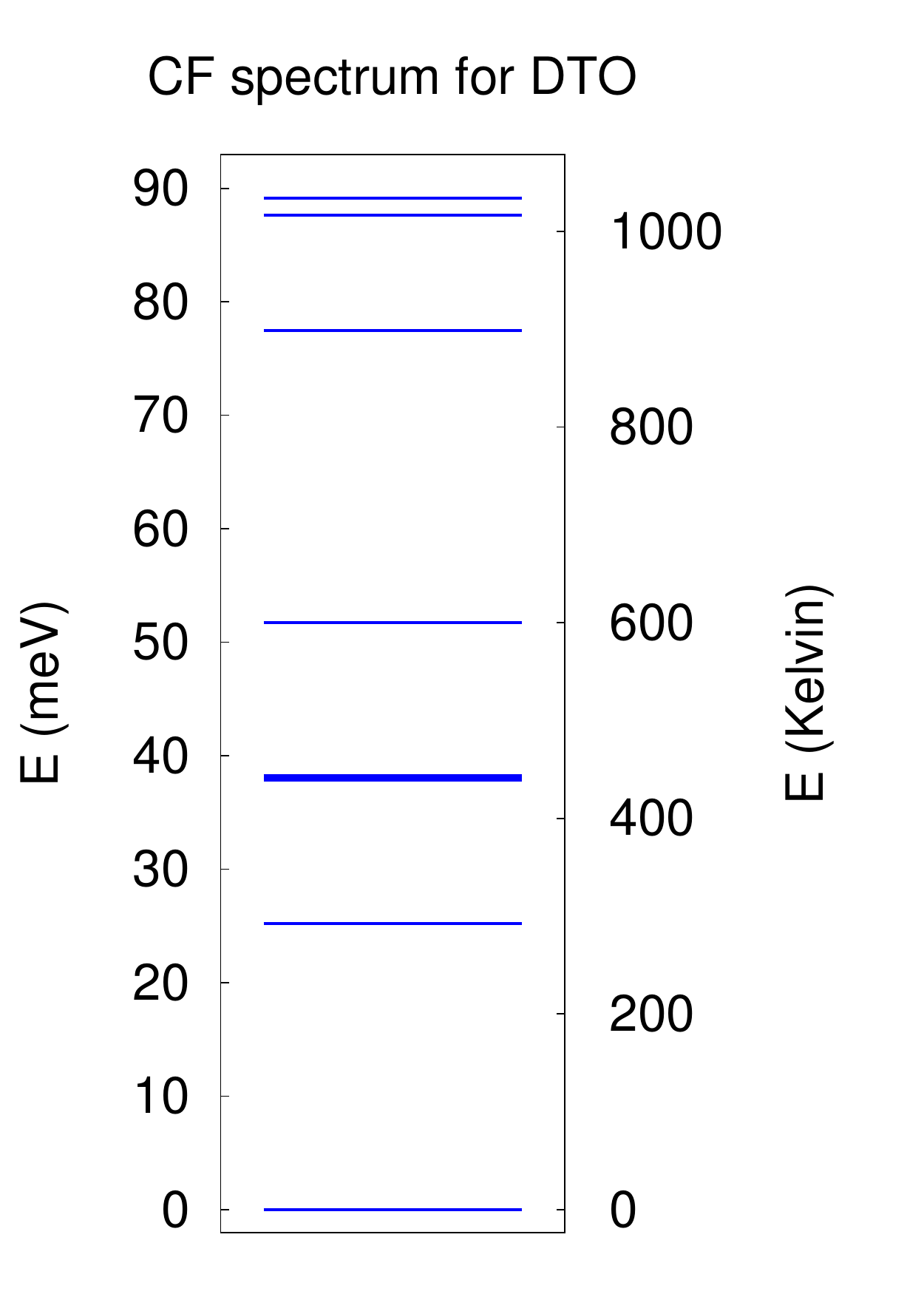}}
	\caption	{
				\label{fig:CFenergies}
				crystal-field spectra for HTO (a) and DTO (b),	respectively.
				The spectrum of HTO features both doublets (solid lines) and singlets (dashed dotted lines).
				In meV, bottom to top, the series of doublets is:  0, 21.96, 25.99, 59.59, 71.51, 76.80 
				while the one of singlets is: 20.42, 27.71, 69.36, 69.94, 80.52.
				In contrast, since Dy$^{3+}$ is a Kramers ion, DTO features only doublets.
				These are eight in total: 0, 25.23, 38.0, 38.21, 51.75, 77.49, 87.65, 89.16. 
				Note the thicker line just below 40~meV is not a quadruplet, but rather it corresponds to the two doublets 38.0, 38.21. 
				}
\end{figure}

The CF Hamiltonian can be diagonalised to obtain the CF states. The spectrum is, in general, made of multiplets and singlets since the Stark splitting, 
induced by the crystalline electric fields, removes only partially the $2J+1$ degeneracy of the ground state multiplet. 
The spectrum of HTO (Fig.~\ref{fig:CFenergiesHTO}) features five singlets and six doublets, 
while the spectrum of DTO (Fig.~\ref{fig:CFenergiesDTO}) is only made of doublets. 
This discrepancy is due to Kramers theorem forbidding singlets in spectra of atoms 
with an odd number of electrons (Ho$^{3+}$ has $n=10$ electrons in the 4-$f$ shell, while Dy$^{3+}$ has $n=9$). 
The order of magnitude for the energies, however, is roughly the same, and the ground state is a doublet in both. The energy gap between the ground state doublet energy and the first excited level is in excess of $200$~K. 

\begin{figure}
	\centering
	\subfloat	[HTO	\label{fig:GSCFmulti_HTO}]
					{\includegraphics[trim = 0mm	0mm		0mm 	0mm,	clip,	width=.9\linewidth]{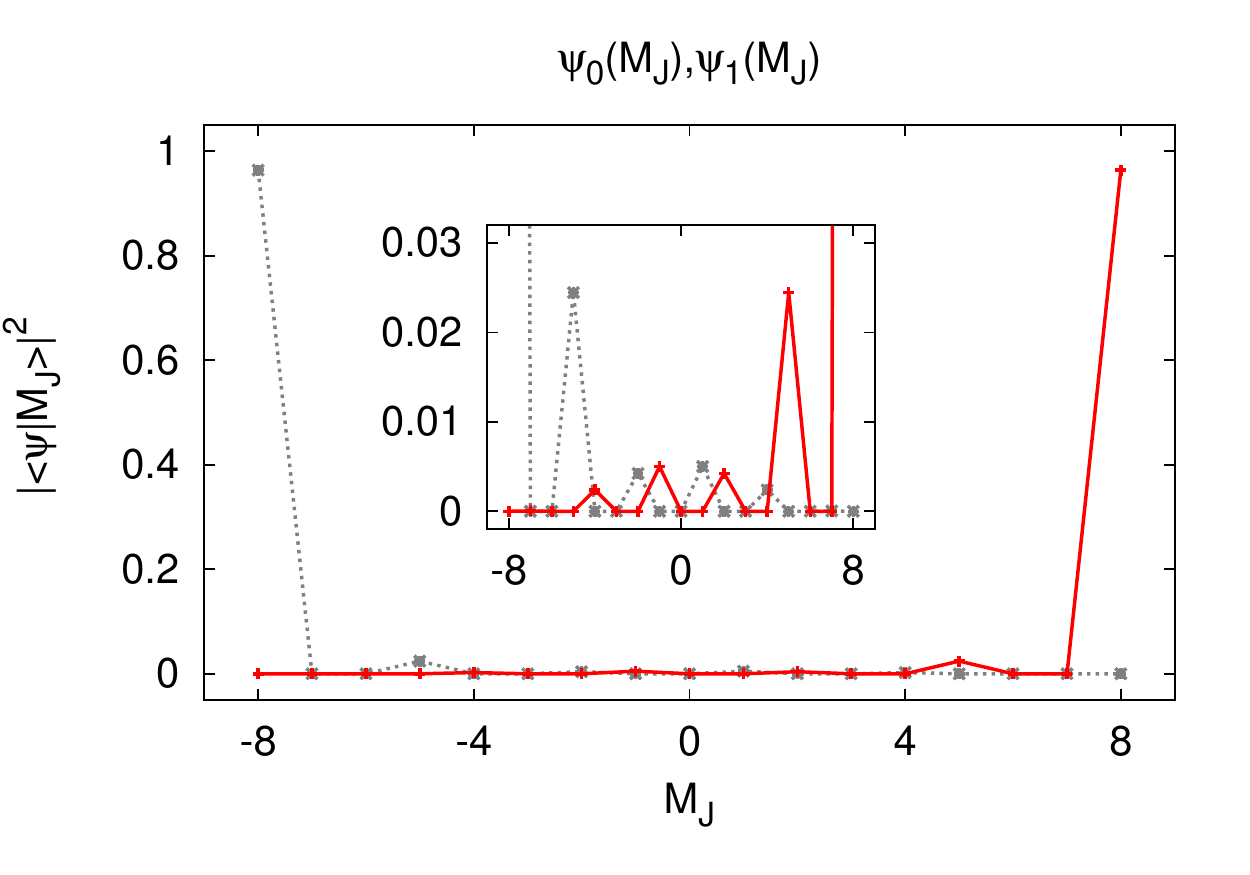}}

	\subfloat	[DTO	\label{fig:GSCFmulti_DTO}]
					{\includegraphics[trim = 0mm	0mm		0mm 	0mm,	clip,	width=.9\linewidth]{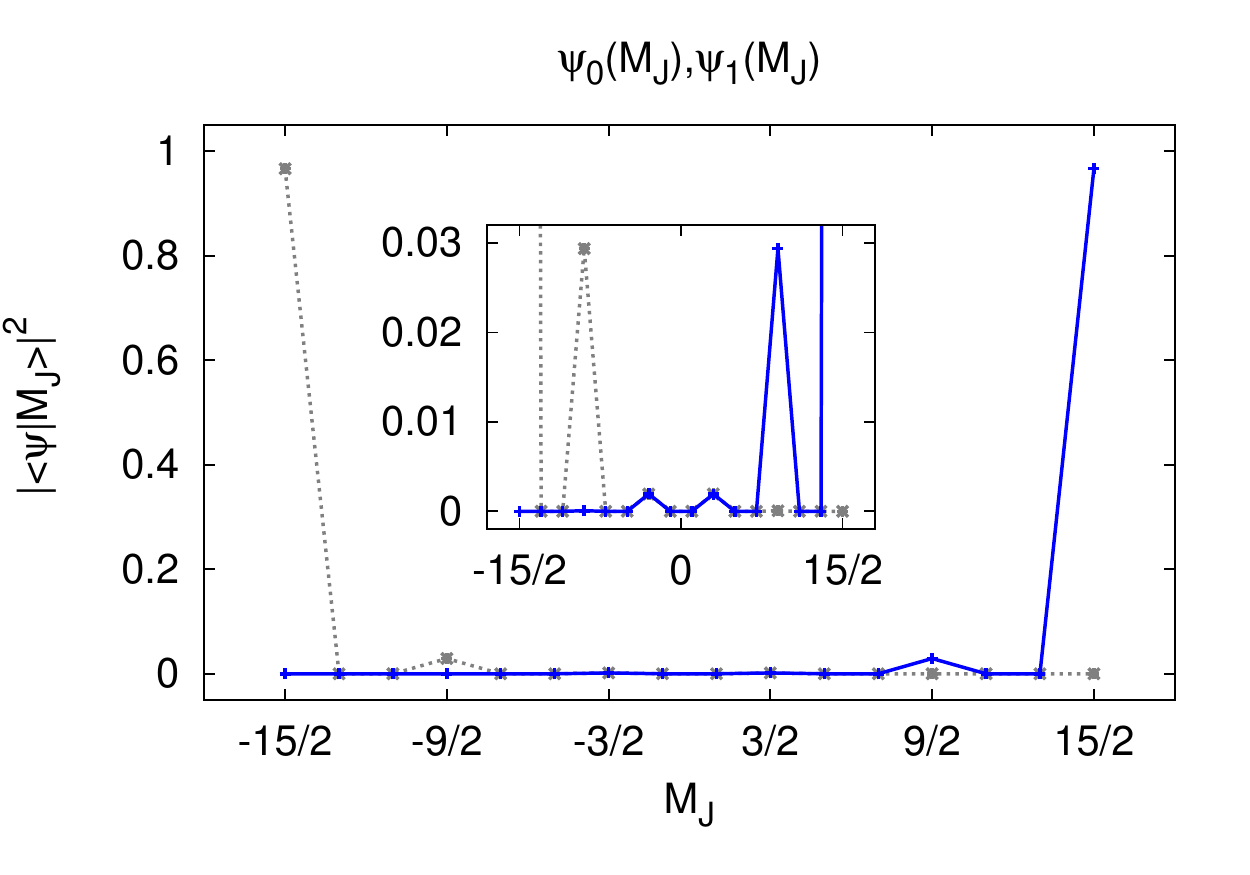}}
	\caption{ 		\label{fig:wave_functions}
						A possible choice of basis wave functions,  $\ket{\psi_{0}}$ (solid lines) and $\ket{\psi_{1}}$ (dotted grey lines), for the ground state doublet of Ho$^{3+}$ (a) and  Dy$^{3+}$ (b). 
						The wave functions have been obtained by diagonalising the crystal-field Hamiltonian in 
						Eq.~\eqref{eq:HamCFpyroStevensShort} with the crystal-field parameters given in Table~\ref{tab:spiniceCFparamStev}.
						Insets: the same shown over a narrower vertical range. 
				}
\end{figure}

Two possible basis eigenfunctions for the ground state doublets are displayed in Fig.~\ref{fig:wave_functions}, showing that they can be well approximated by the fully polarised states 
$ \ket{ \psi_{0}}	\approx \ket{ M_{J}= J}, \ket{ \psi_{1}} \approx	 \ket{M_{J}=- J}$. 
This illustrates the strong anisotropy along the local quantisation axis in both systems. 

\section	{	Effect of a magnetic field		\label{sec:MagnField}		}

The degeneracy of the crystal-field spectra is removed in the presence of a magnetic field $\mathbf{B}$:
\begin{equation}
	\hat{\mathcal{H}}	=	\hat{\mathcal{H}}_{\mathrm{CF}}	-	g_{J}		\mu_{\mathrm{B}}	\,	\hat{\mathbf{J}}	\cdot		\mathbf{B}.
\label{eq:Ham_Cf_Zeeman}
\end{equation}
In this equation, $ \mu_{\mathrm{B}}= e \hbar /2 m_{e}$ is the Bohr magneton 
($e$ and $ m_{e}$ are, respectively, the charge and the mass of the electron) 
and $g_{J}$ is the Land\'e factor for the RE$^{3+}$ ion with total angular momentum $\hat{\mathbf{J}}$  
($g_{J}=5/4$ and $g_{J}=4/3$, respectively, for Ho$^{3+}$ and Dy$^{3+}$). 

In the following, we use the local coordinate system 
\begin{equation}	\label{eq:LocCoo0alone}
			\mathbf{x}_{0}		=		\frac{1}{\sqrt{6}}(1,1,-2)	,	\;
			\mathbf{y}_{0}		=		\frac{1}{\sqrt{2}}(-1,1,0)	,	\;	 
			\mathbf{z}_{0}		=		\frac{1}{\sqrt{3}}(1,1,1)	,
\end{equation}
with respect to the global axes $\mathbf{X,Y,Z} $ of the cubic pyrochlore unit cell (see Fig.~\ref{fig:CFandBtop}), with the $\mathbf{z}_{0}$ axis conveniently pointing along the high-symmetry direction of the crystal-field Hamiltonian~\cite{Rosenkranz:2000,Malkin:2010}. 

\subsection{	Exact diagonalisation 	\label{sec:MagnField_ExactDiag}	}

A longitudinal field along the local easy axis leads to conventional Zeeman splitting linear in field strength and selects of one of the two polarised states in Fig.~\ref{fig:wave_functions}. At similar field strengths, this is the field direction that results in the largest energy splitting due to the anisotropy. 

\begin{figure}
	\centering
	\includegraphics[trim = 0mm 0mm 0mm 0mm, clip, width=1\linewidth]{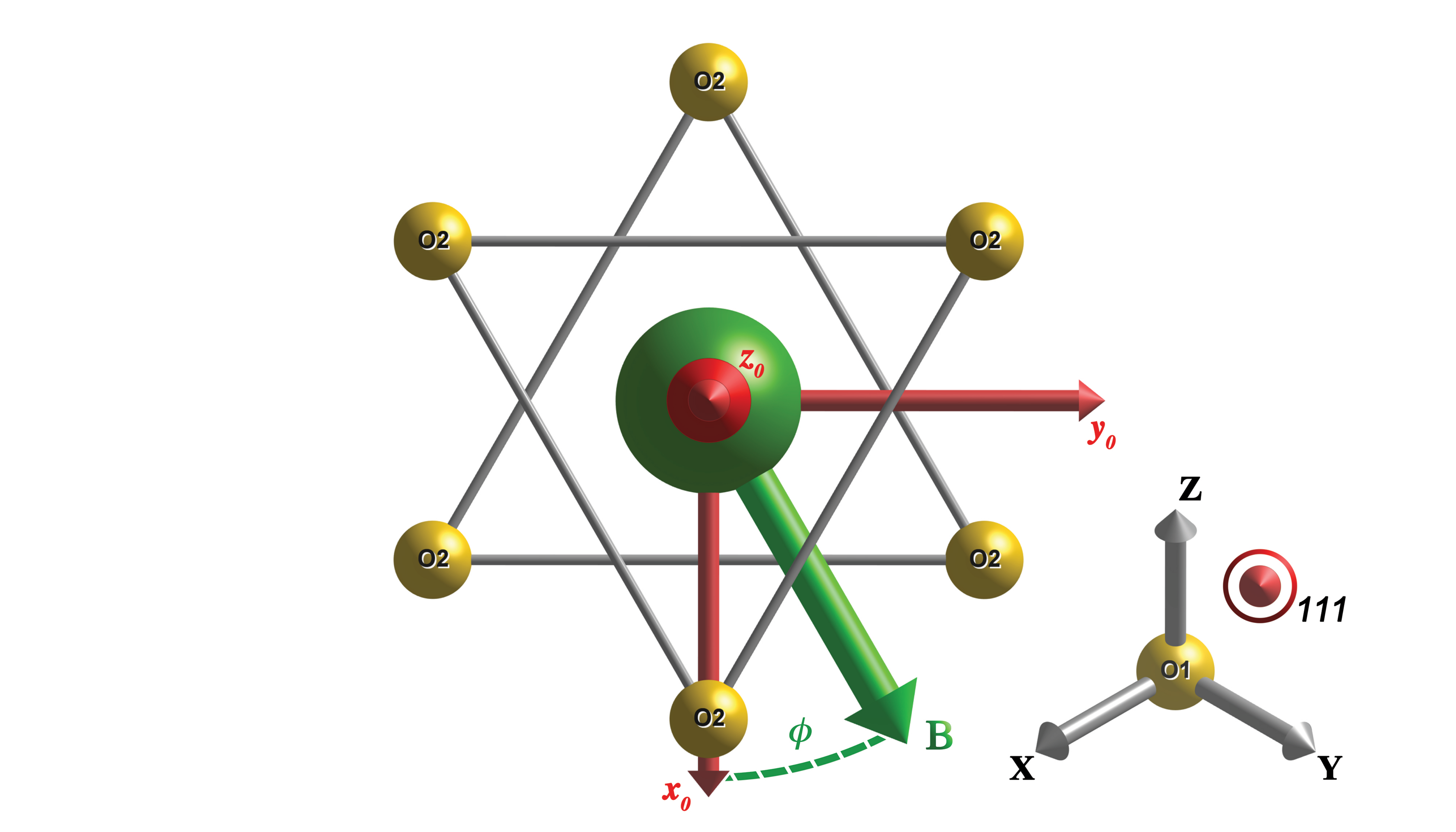}
	\caption{	\label{fig:CFandBtop}
                                        The local coordinate frame
                                        $\mathbf{x}_{0},\mathbf{y}_{0},\mathbf{z}_{0}$
                                        (red arrows)
                                        used to describe the transverse magnetic field                                                  $\mathbf{B}$ (green arrow) and its direction angle $\phi$
                                        in Eq.~\eqref{eq:perturbation}. 
                                      The atoms are shown in the same top-view as in Fig.~\ref{fig:CFenvironment(c)}.
				}
\end{figure}

If the longitudinal component vanishes and a purely transverse field component is present, then the chosen polarised basis states split into symmetric ``bonding'' and antisymmetric ``anti-bonding'' combinations. A polarization in the plane perpendicular to the easy axis is however opposed by the anisotropy and this competition results in unusual effects that will be discussed in the following. 

The coupling of the total angular momentum to the transverse magnetic field can be written in terms of ladder operators as 
\begin{equation}
	\hat{\mathbf{J}}		\cdot		\mathbf{B_{\bot}}		=	
	\frac{1}{2}	\left|	\mathbf{B_{\bot}}		\right|	\left(	e^{-i\phi}	\hat{J}_{+}		+	e^{i\phi}	\hat{J}_{-}		\right) 
\,	
. 
\label{eq:perturbation}
\end{equation}
In the local coordinate system, 
$\phi$ is the angle of the field with respect to $\mathbf{x}_{0}$ in the plane  transverse to the easy axis $\mathbf{z}_{0}$  (see Fig.~\ref{fig:CFandBtop}). 

\begin{figure}[ht]
	\centering
	\includegraphics[trim = 0mm 0mm 0mm 0mm, clip, width=\linewidth]{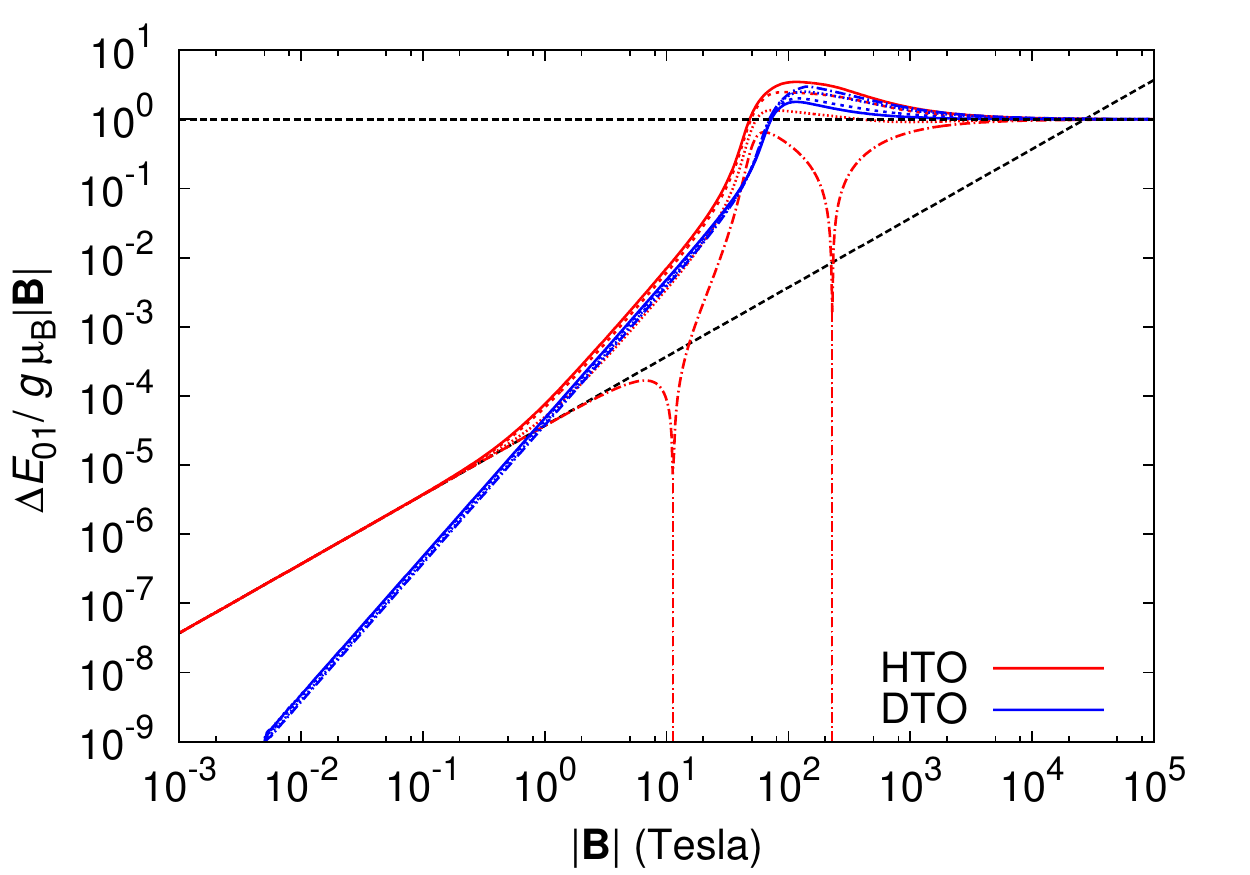}
	\caption	{	
				\label{fig:Splitting}
				Splitting of the ground state doublet under 
				the influence of a purely transverse magnetic field. 
				The red curves correspond to the non-Kramers behaviour of HTO, 
				while the blue curves correspond to the Kramers behaviour of DTO.
				Note the y axis is dimensionless to allow a consistent comparison of the two systems. 
				The different curves correspond to fixed angles of the transverse field:
				$\phi=0\,^{\circ}$ (solid curve), $\phi=10\,^{\circ},$ (short-dashed), $\phi=20\,^{\circ}$ (dotted) and $\phi=30\,^{\circ}$ (dotted-dashed);
				angles of $ \phi + n \, 120 ^{\circ} $, with $n$ integer, give exactly the same curves because of the CF trigonal symmetry of the O2 ions.
				The two long-dashed straight lines show the limiting behaviours 
				at very high fields -- Larmor precession, Eq.~\eqref{eq:SplittingLarmor} -- and at low fields for HTO -- degenerate perturbation theory, Eqs.~(\ref{eq:power_law_Ho3+}) and (\ref{eq:slopeHo}).
			}
\end{figure}

The dependence of the splitting of the ground state doublet $ \Delta E_{01} $ vs the magnitude of the transverse field is shown in Fig.~\ref{fig:Splitting} for both HTO and DTO. 
For very large fields the anisotropic effect of the CF environment becomes negligible and the magnetic moments undergo simple Larmor precession with frequency $ \omega_{\mathrm{L}} $ given by 
\begin{equation}
	\Delta E_{01}	=	\hbar		\omega_{\mathrm{L}}	=	g_{J}		\mu_{\mathrm{B}}	\left|	\mathbf{B}		\right|	\,	.
	\label{eq:SplittingLarmor}
\end{equation}
Due to the strong crystal-fields in HTO and DTO, 
such regime is clearly experimentally unattainable ($B > 10^3$~T). 
This illustrates the strength of the energy scales set by the crystal-field 
and provides a reference for magnetic field values that can be considered a small perturbation. 

\begin{figure}[ht!]
	\centering
	\subfloat	[HTO	\label{fig:DeltaEvsAngleHTO}]
					{\includegraphics[trim = 0mm	0mm	0mm 	0mm,	clip,	height=0.6\columnwidth]{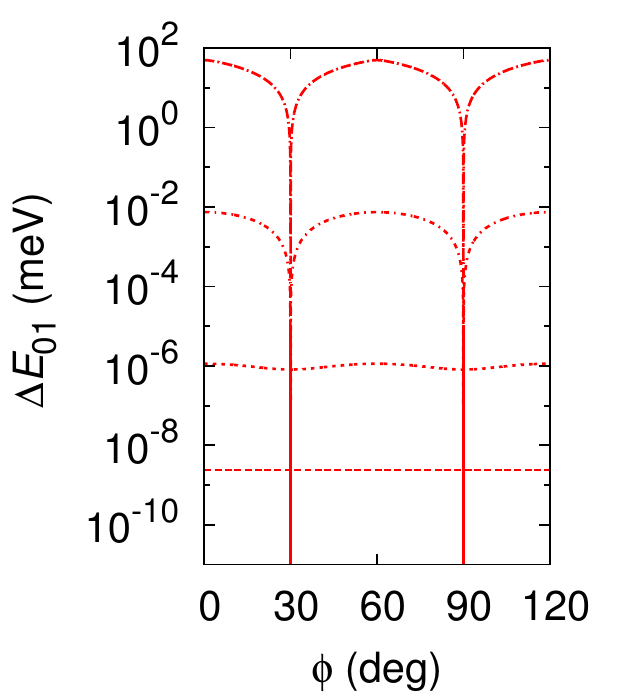}}					
	\subfloat	[DTO	\label{fig:DeltaEvsAngleDTO}]
					{\includegraphics[trim = 9mm	0mm		0mm 	0mm,	clip,	height=0.6\columnwidth]{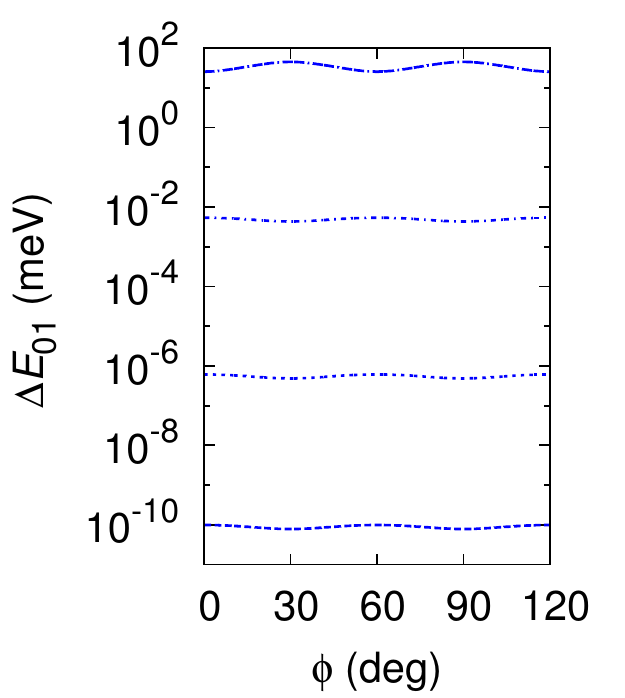}}		
	\caption	{	
				\label{fig:DeltaEvsAngle}
				Ground state splitting for both HTO (a) and DTO (b), 
				as a function of the angle $\phi$ at four particular values of the transverse field (from bottom to top in each figure):
				$B=0.03$~T, $B=0.55$~T, 
 				$B=11.31$~T, and $B=228.33$~T. 
				For HTO, the two strongest field values resonate and 
				make the splitting close at $\phi= 30 ^{\circ} + n \, 60 ^{\circ}$ ($n$ integer). 
			}
\end{figure}

At lower fields, when the two competing terms in Eq.~\eqref{eq:Ham_Cf_Zeeman} have comparable energies, the response of the system becomes anisotropic. 
This anisotropy is much stronger for Ho$^{3+}$ than for Dy$^{3+}$ and, for $\phi= 30 ^{\circ} + n \, 60 ^{\circ}$ with $n$ integer,
it leads to resonances (due to level crossing between $E_{0}$ and $E_{1}$)
shown in Fig.~\ref{fig:Splitting} (red dotted-dashed line) and in Fig.~\ref{fig:DeltaEvsAngleHTO}. 

Finally, at fields of the order of 1~T or less, the ion enters a perturbative
regime where $\Delta E_{01}$ is given by the following power-laws: 
\begin{subequations}
	\begin{alignat}{2}
		\Delta E_{01} & =	\;	\alpha_{\mathrm{HTO}}^{(2)}			
							\left|\mathbf{B}\right|^{2}
						&	\qquad	\text{for HTO},			\label{eq:power_law_Ho3+}	\\[0.2cm]
		\Delta E_{01} & =	\;	\alpha_{\mathrm{DTO}}^{(3)}	(\phi)	
							\left|\mathbf{B}\right|^{3}
						&	\qquad	\text{for DTO}.			\label{eq:power_law_Dy3+}
	\end{alignat}
	\label{eq:PowerLaws}
\end{subequations}
Ho$^{3+}$ is not a Kramers ion and features some singlets in its unperturbed energy spectrum. 
These are responsible for the quadratic behaviour (low field asymptotics in Fig.~\ref{fig:Splitting}), whose coefficient 
\begin{equation}	
\alpha_{\mathrm{HTO}}^{(2)}=2.68\times10^{-6}\frac{ \mbox{meV}}{ \mbox{T}^{2}} 
\label{eq:slopeHo}
\end{equation}
can be obtained analytically from perturbation theory (see Sec.~\ref{sec:MagnField_PerturbTheory}). 
Dy$^{3+}$ instead is a Kramers ion and all unperturbed energy levels are doublets. As we explain in the next section, this causes the quadratic correction to vanish identically, leading to a cubic dependence on the applied field.  
Fitting the corresponding asymptotic low-field behaviour in Fig.~\ref{fig:Splitting}, we obtain the angle-dependent coefficient 
\begin{equation}	
		\alpha_{\mathrm{DTO}}^{(3)} (\phi)	=	
		6.8\times10^{-7}	\big(	\,	1+A \cos	\left(	6 \phi \right)	\big)	\frac{ \mbox{meV}}{ \mbox{T}^{3}}
	\label{eq:slopeDy}
\end{equation}
with $A=0.114$.

\subsection{	Perturbation theory	\label{sec:MagnField_PerturbTheory}	}

We can gain insight into the low field behaviour by using (degenerate) perturbation theory on 
\begin{equation}
	\hat{\mathcal{H}}	=	\hat{\mathcal{H}}_{0}		-	\lambda\hat{V} 
	\,	, 
	\label{eq:HamPerturb}	
\end{equation}
where $\hat{\mathcal{H}}_{0}	\equiv	\hat{\mathcal{H}}_{\mathrm{CF}}$ is the CF Hamiltonian in Eq.~\eqref{eq:HamCFpyroStevensShort} and the perturbation 
$\hat{V}\equiv\mathbf{\mathfrak{\mathcal{E}}_{\mathrm{CF}}} \, \hat{\mathbf{J}}\cdot\mathbf{B}/\mathbf{|B|}$ corresponds to the Zeeman energy in Eq.~\eqref{eq:perturbation}, 
tuned by the dimensionless parameter  $\lambda=g_{J}\mu_{\mathrm{B}}\left|\mathbf{B}\right|/\mathbf{\mathfrak{\mathcal{E}}_{\mathrm{CF}}}$,
where $\mathbf{\mathfrak{\mathcal{E}}_{\mathrm{CF}}}$ is an arbitrary reference energy scale, e.g., related to the CF bandwidth.
It is useful to introduce $\ket{\psi_{n}^{(0)}}$ as the (unperturbed) CF eigenstates with energy $E_{n}^{(0)}$ ($n=0,...,2J$). 

The splitting of the RE$^{3+}$ ground state doublet is given by
\begin{equation}
	\begin{split}
		& \Delta E_{01} 		=  	\lambda\sqrt{(V_{0,0}-V_{1,1})^{2}+4\left|V_{0,1}\right|^{2}}
						\\
		&				+	\lambda^{2}	\sqrt{
											\left(
												\sum_{k>1}\frac{
																\left|V_{0,k}\right|^{2}-\left|V_{1,k}\right|^{2}
																}{
																\Delta E_{0k}^{(0)}
																}
											\right)^{2}
											+4\left|	\sum_{k>1}\frac{V_{0,k}V_{k,1}}{\Delta E_{0k}^{(0)}}	\right|^{2}
											}
	\end{split}							   
	\label{eq:Delta01Perturb}
\end{equation}
up to second order in $\lambda$. 
In this expression $V_{n,m} \equiv\bra{\psi_{n}^{(0)}}\hat{V}\ket{\psi_{m}^{(0)}}$ and $\Delta E_{0k}^{(0)}=E_{k}^{(0)}-E_{0}^{(0)}$. 

Firstly, we notice that both HTO and DTO have $V_{0,0}=V_{1,1}=V_{0,1}=0$ (see App.~\ref{sec:App_MatElemPerturb}), and therefore the first order contribution vanishes identically. 

In order to evaluate the second order contribution, we need to consider matrix elements of the transverse field perturbation $\hat{V}$ between the ground state doublet and the excited states. The symmetries of these matrix elements reflect the symmetries of the crystal-field environment. 

In App.~\ref{sec:App_MatElemPerturb} we discuss the different contributions in detail. We find two different behaviours for the excited states that form doublets. Some of them (type A) have identically vanishing matrix elements with the ground state doublet, $V_{0,n} = V_{1,n} = 0$ for $n$ belonging to type A, and they trivially do not contribute to the splitting at second order. The other doublets (type B) have non-vanishing matrix elements which satisfy the following relations: 
\begin{eqnarray}
\left\vert V_{0,m}\right\vert^{2} = \left\vert V_{1,m+1}\right\vert^{2}
\label{eqA:VD0-1-1 a}
\\ 
\left\vert V_{1,m}\right\vert^{2} = \left\vert V_{0,m+1}\right\vert^{2} 
\label{eqA:VD0-1-1 b}
\\
V_{0,m} V_{m,1} + V_{0,m+1} V_{m+1,1}	=	0
, 
\label{eqA:VD0-1-1 c}
\label{eqA:VD0-1-1}
\end{eqnarray}
where $\ket{\psi_{m}^{(0)}}$ and $\ket{\psi_{m+1}^{(0)}}$ are the two eigenstates belonging to an excited doublet of type B 
(see Eq.~\eqref{eqA:VDD:10:1HTO} and Eq.~\eqref{eqA:VDD:10:1DTO} in App.~\ref{sec:App_MatElemPerturb}). 
These relations imply that also doublets of type B do not contribute to the splitting at second order.  
(Details of the A- and B-type doublet wavefunctions and their matrix elements with the transverse field operator are given in Tables \ref{tab:HTOstates}-\ref{tab:DTOstates} in App.~\ref{sec:App_MatElemPerturb}.) 

These results hold for both DTO and HTO. 
The former only has doublets (of either type A or B) in the spectrum due to Kramers degeneracy and no splitting occurs at second order. Indeed, the second order term in Eq.~\eqref{eq:Delta01Perturb} generally reduces to the sum of the contributions from the singlets alone (see Eq.~\eqref{eqA:Vs0} in App.~\ref{sec:App_MatElemPerturb}). Fig.~\ref{fig:Splitting} clearly shows that a non-vanishing third order contribution does exist, which we extract by fitting, Eq.~\eqref{eq:slopeDy}. 

HTO, on the contrary, has some singlets amongst its excited states, which give a non-vanishing second-order contribution to the splitting. This can be readily computed, Eq.~\eqref{eq:slopeHo}, and is in excellent agreement with the slope found from the numerical simulations
(see the corresponding long-dashed straight line in Fig.~\ref{fig:Splitting}). 
We notice that the third order contribution has an angular dependence on $\phi$ which is absent at second order. 

It is interesting to notice in Fig.~\ref{fig:Splitting} that the cubic power-law found for DTO persists up to $\sim 100$~T. 
In contrast, the quadratic power law in HTO begins to break at fields of the order of $0.1$~T, depending on the in-plane angle $\phi$, holding up to almost $10$~T for $\phi = 30 \,^{\circ}$; 
for all other angles, the cubic term becomes clearly dominant in the range from $1$~T to $10$~T, with an angular dependence similar to the one for DTO.

\subsection{Doublet splitting and time scales 		\label{sec:timescales}}

Let us compare the observed ground state doublet splitting in Fig.~\ref{fig:Splitting} with experimental magnetic relaxation time scales in spin ice~\cite{Snyder:2004}. The latter are typically of the order of $1$~ms (at least in DTO), which corresponds to an approximate energy splitting of $10^{-7}-10^{-8}$~K. 

In order to estimate the former, one needs typical values for the exchange and dipolar transverse field strength. Ref.~\onlinecite{Sala:2012} suggests the range $0.1-1$~T. Using Eqs.~\eqref{eq:power_law_Ho3+},~\eqref{eq:power_law_Dy3+},~\eqref{eq:slopeHo} and \eqref{eq:slopeDy}, we find that this corresponds to splittings in the range of $10^{-8}$~K to $10^{-5}$~K. 

This rough theoretical estimate is consistent with the experimental value. 
Whilst further investigation is clearly needed, the result is nonetheless suggestive that internal fields generated by exchange and dipolar interactions can in principle be responsible for (single-ion) quantum spin-flip dynamics in spin ice.

\subsection{	Anisotropic response to a transverse field 	 	\label{sec:MagnField_AnisBehav}	}

\begin{figure*}
	\centering
	\subfloat	[ $\braket{\hat{J}_{x}}$ - $T=0$ K - HTO	\label{fig:MagnetizAngleTeslaMxHo}]
					{\includegraphics[trim = 0mm	7mm	0mm 	12mm,	clip,	width=.48\linewidth]{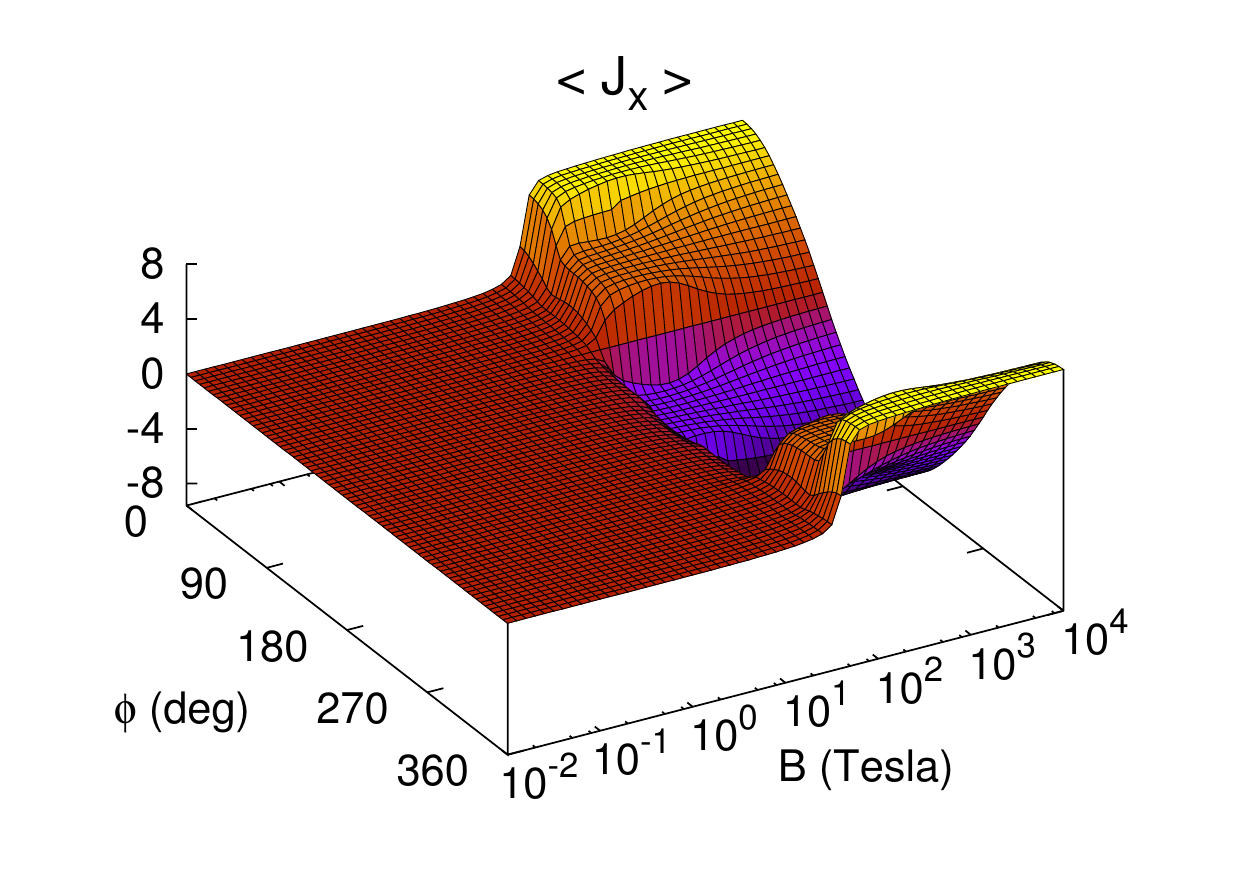}}
	\subfloat	[ $\braket{\hat{J}_{x}}$ - $T=0$ K - DTO	\label{fig:MagnetizAngleTeslaMxDy}]
					{\includegraphics[trim = 0mm	7mm	0mm 	12mm,	clip,	width=.48\linewidth]{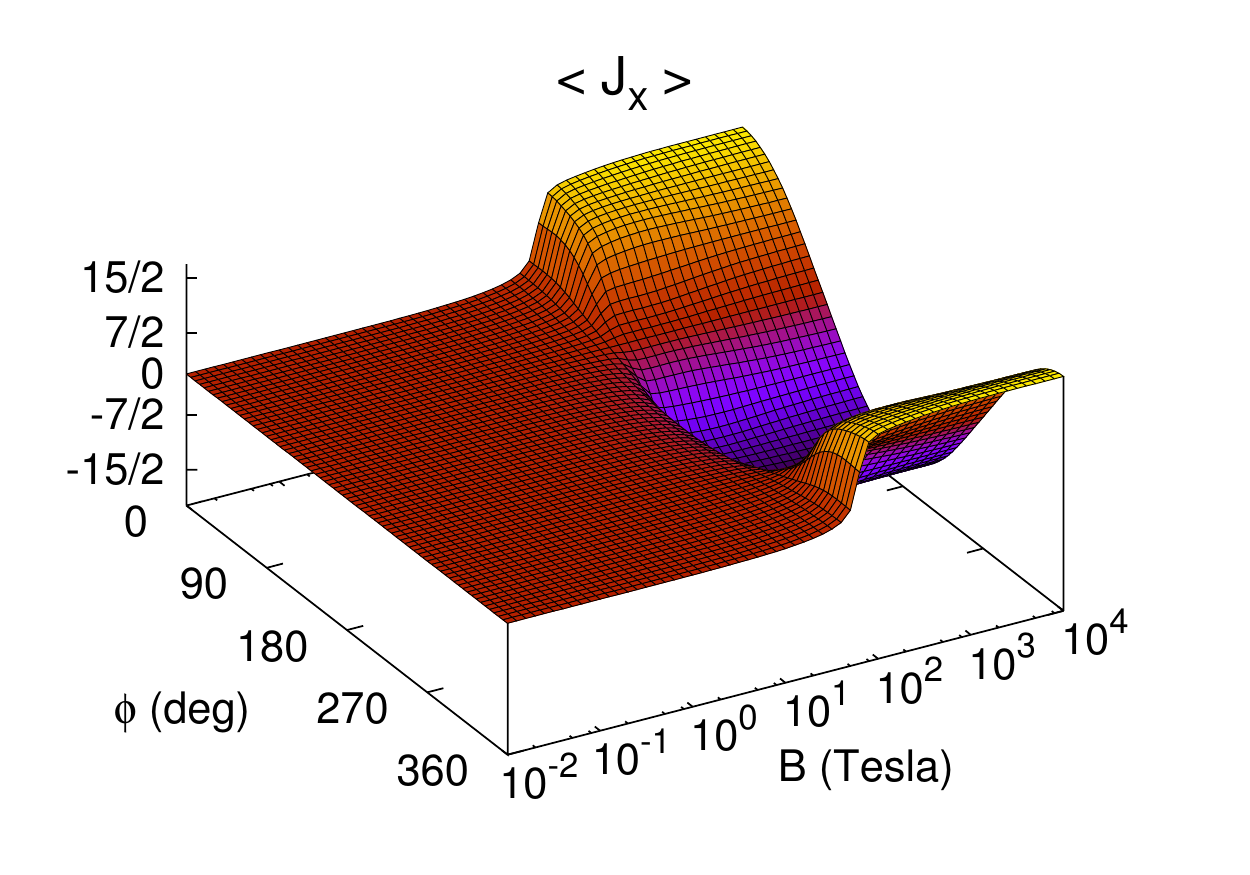}}

\vspace{-4mm}

	\subfloat	[ $\braket{\hat{J}_{y}}$ - $T=0$ K - HTO	\label{fig:MagnetizAngleTeslaMyHo}]
					{\includegraphics[trim = 0mm	7mm	0mm 	13mm,	clip,	width=.48\linewidth]{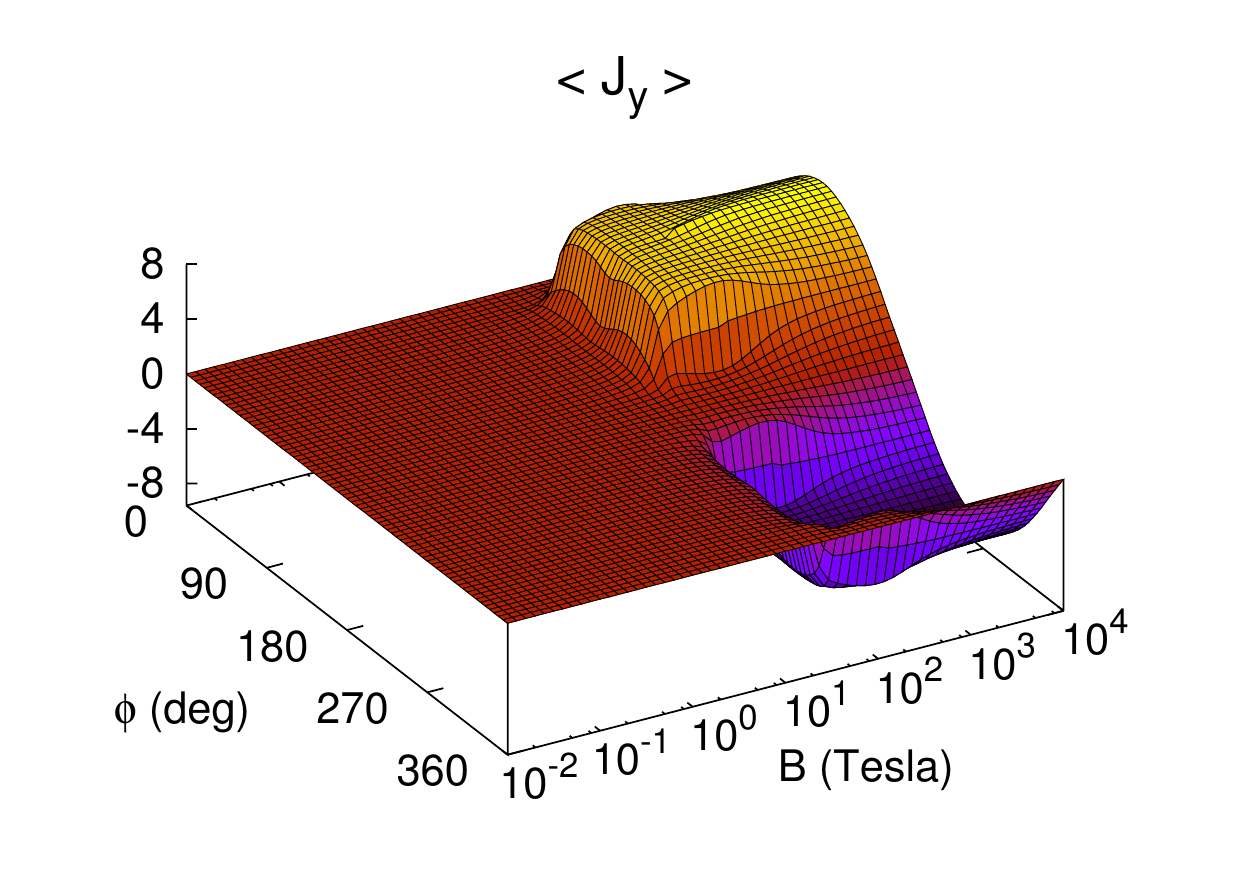}}
	\subfloat	[ $\braket{\hat{J}_{y}}$ - $T=0$ K - DTO	\label{fig:MagnetizAngleTeslaMyDy}]
					{\includegraphics[trim = 0mm	7mm	0mm 	13mm,	clip,	width=.48\linewidth]{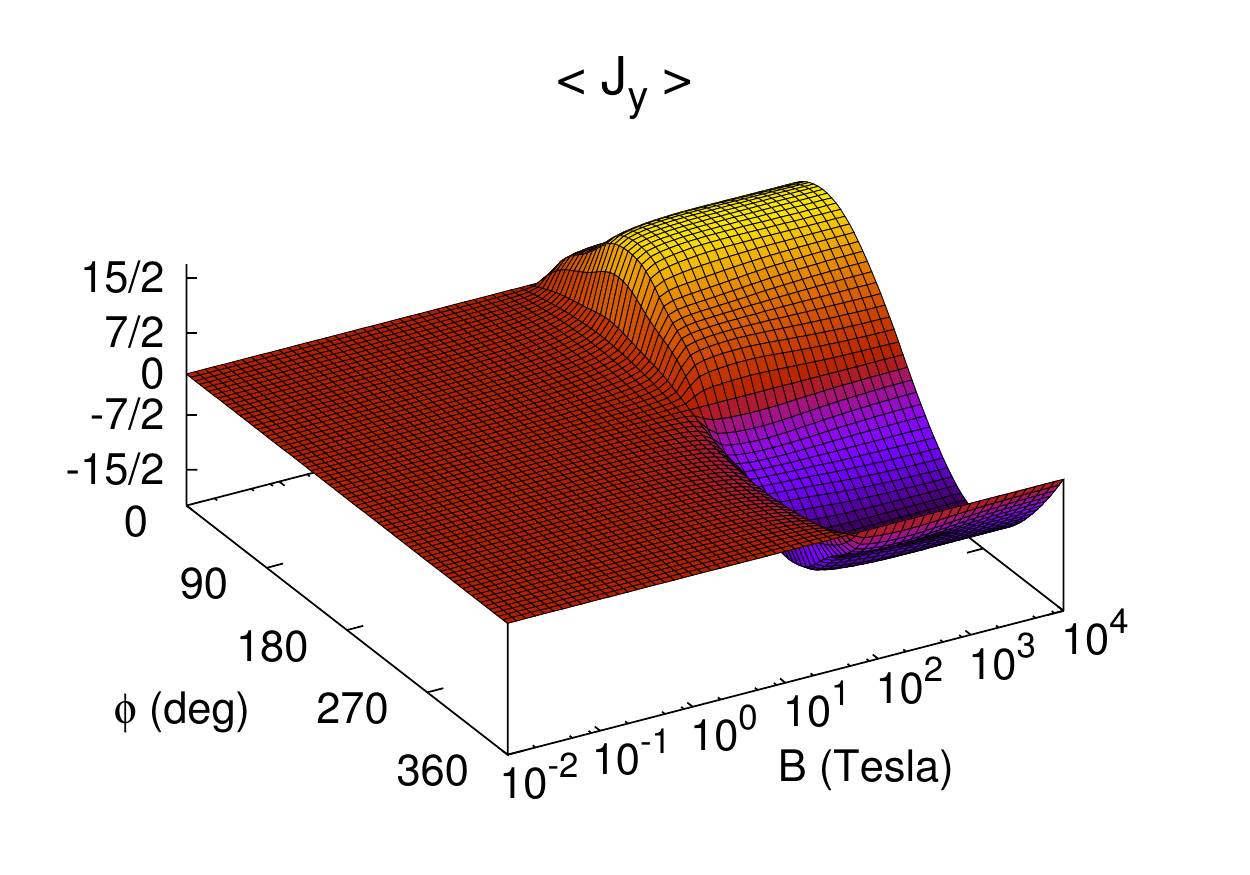}}

\vspace{-4mm}

	\subfloat	[ $\braket{\hat{J}_{z}}$ - $T=0$ K - HTO	\label{fig:MagnetizAngleTeslaMzHo}]
					{\includegraphics[trim = 0mm	7mm	0mm 	14mm,	clip,	width=.48\linewidth]{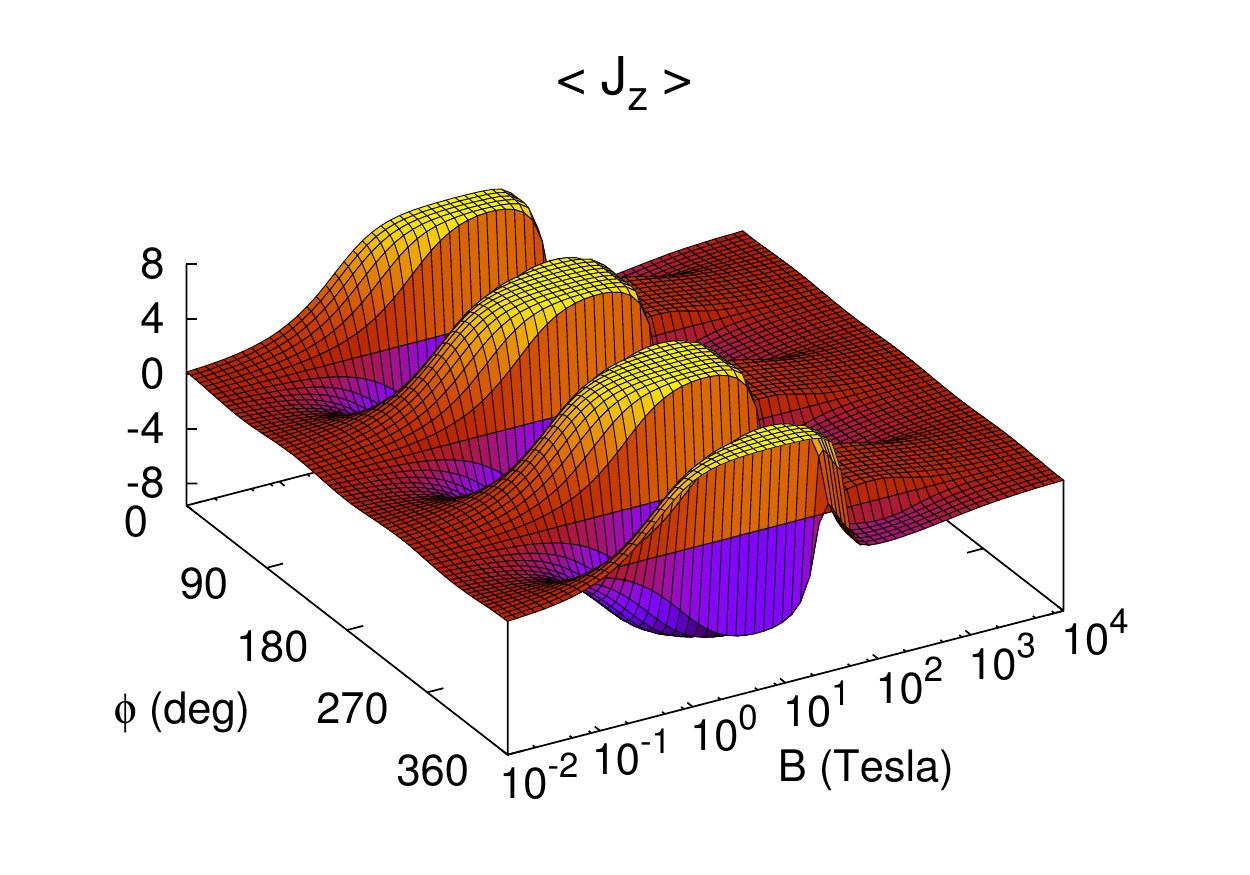}}
	\subfloat	[ $\braket{\hat{J}_{z}}$ - $T=0$ K - DTO	\label{fig:MagnetizAngleTeslaMzDy}]
					{\includegraphics[trim = 0mm	7mm	0mm 	14mm,	clip,	width=.48\linewidth]{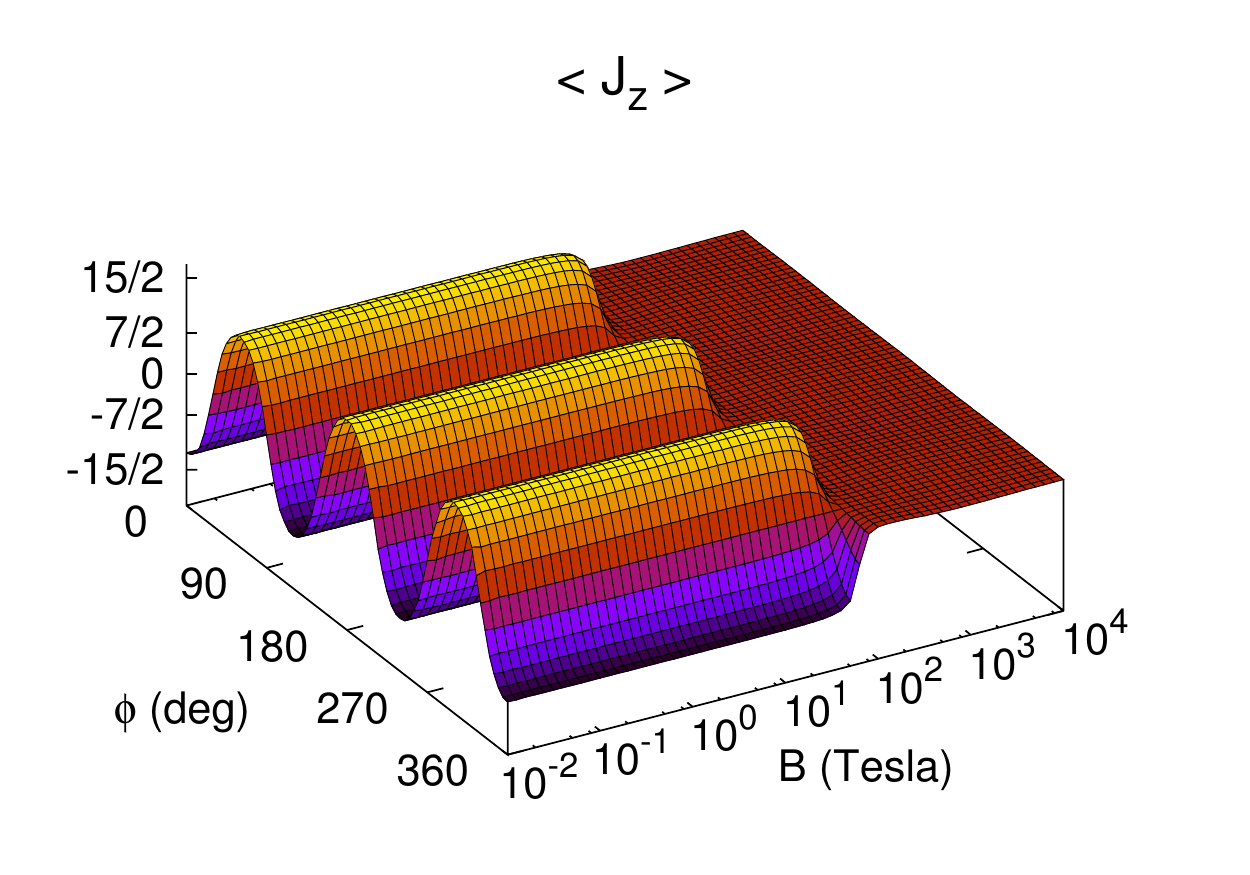}}

\vspace{-4mm}

	\subfloat	[ $\braket{\hat{J}_{z}}$ - $T=500$ mK - HTO 	\label{fig:JzTempAngleTeslaz500mK_Ho}]
					{\includegraphics[trim = 0mm	7mm	0mm 	14mm,	clip,	width=.48\linewidth]{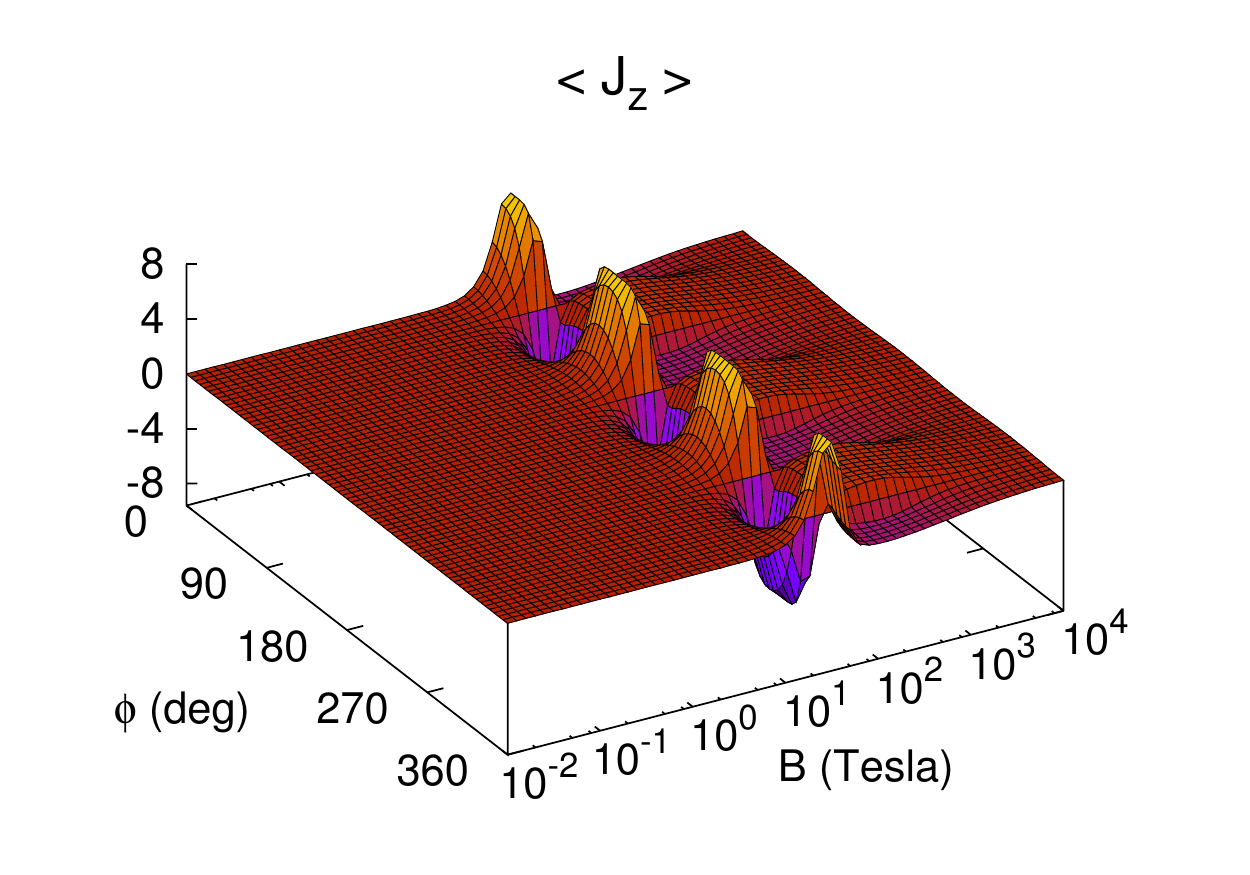}}
	\subfloat	[ $\braket{\hat{J}_{z}}$ - $T=500$ mK - DTO	\label{fig:JzTempAngleTeslaz500mK_Dy}]
					{\includegraphics[trim = 0mm	7mm	0mm 	14mm,	clip,	width=.48\linewidth]{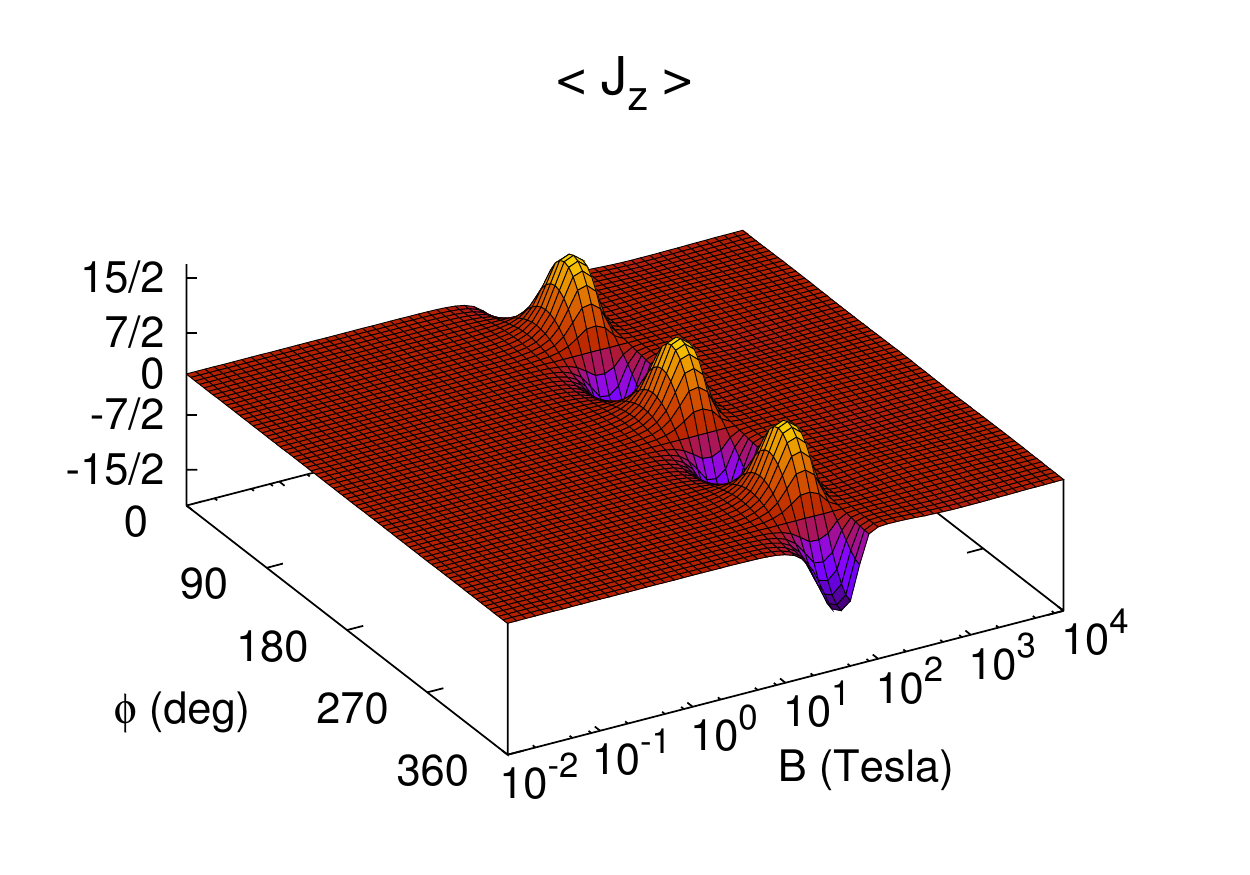}}
	\caption	{
				\label{fig:MagnetizAngleTesla}
				\label{fig:JzTempAngleTesla}
        Panels (a-f): 
				Expectation values for the three components of the total angular momentum $\hat{\mathbf{J}}$
				in the ground state of the Hamiltonian in Eq.~\eqref{eq:Ham_Cf_Zeeman} 
				with purely transverse magnetic field: 
				$\braket{\hat{J}_{\alpha}}=  \braket{ \psi | \hat{J}_{\alpha} | \psi }$,
				$\alpha=x,y,z$, 
				as function of the angle $\phi$ and the strength $|\mathbf{B}|$ of the field in logarithmic scale. 
				For both HTO (left) and DTO (right), the $x,y$ components are negligible for fields up to $10$~T. 
				In contrast, the $m_{z}$ components feature a sizeable periodic dependence on the angle $\phi$ below $10$~T. 
				This is a manifestation of the strong axial anisotropy characterising the ground state of the spin ice RE$^{3+}$ ions.
				Note the different response in the two systems: 
				DTO features a smooth angular dependence which becomes asymptotically constant in the low field limit 
				(from $10$~T down to the lowest fields), 
				whereas the oscillatory behaviour in HTO is more abrupt and its amplitude descreases from the saturated value reached at approximately $10$~T, down to zero at low fields. 
				Panels (g-h): 
				Finite temperature behaviour of the expectation value $\braket{\hat{J}_{z}} = \mathrm{Tr}( {\hat{J}_{\alpha} \hat{\rho}} )/\mathrm{Tr}( \hat{\rho} )$ at $T = 0.5$~K for HTO (left) and DTO (right). 
				The Boltzmann weights from the density operator average the two (lowest-energy) states with opposite polarisation along $\mathbf{z}_{0}$.
			}
\end{figure*}

The strong single-ion anisotropy plays a crucial role also in the magnetostatic behaviour of the RE$^{3+}$ ions 
at zero temperature. This is illustrated in Fig.~\ref{fig:MagnetizAngleTesla}, panels a-f, as a function of the angle $\phi$ and of the field strength $| \mathbf{B}|$, 
where 
$ \braket{\hat{J}_{\alpha}}=  \braket{ \psi | \hat{J}_{\alpha} | \psi } $, 
and $ \alpha = x, y, z $ label the three components for the local coordinate system in Fig.~\ref{fig:CFandBtop}. 

Both HTO and DTO acquire negligibly small values of $\braket{\hat{J}_{x}}$ and $\braket{\hat{J}_{y}}$ for fields up to $10$~T (see also Sec.~\ref{sec:FiniteTempMagnMom}),. 
Moreover, we observe a sizeable (zero temperature) response in $ \braket{\hat{J}_{z}} $ to a purely transverse field, signalling non vanishing off-diagonal components of the $g$-tensor. 
Of course, for high enough fields ($|\mathbf{B}| \gg 10$~T), all expectation values tend to the angular dependence of the Larmor regime, as expected when the Zeeman energy dominates over the CF Hamiltonian. 

The main difference between HTO and DTO is in the behaviour of $ \braket{\hat{J}_{z}} $ below $10$~T. 
In HTO, for fields $1 \, \mbox{T} \lesssim | \mathbf{B}| \lesssim 10 \, \mbox{T}$, $\braket{\hat{J}_{z}}$ oscillates rather abruptly with respect to the angle $\phi$ between the saturated values $-8$ and $8$; 
for fields below $1$~T the amplitude of oscillation decreases and it becomes vanishingly small at low fields. 
On the contrary, in DTO the angular dependence is smoother and approaches constant (maximum) amplitude for fields below 10 Tesla; 
the amplitude however never reaches saturation. 
The period of oscillations is the same in both HTO and DTO ($ 120 \,^{\circ} $), but we observe a phase difference of $ 60 \,^{\circ} $. 

We notice that the behaviour of $\braket{\hat{J}_{z}}$ in Fig.~\ref{fig:MagnetizAngleTesla} is consistent 
with the arrangement of the oxygens surrounding the rare earth ions. 
Depending on the $\phi$ angle in the $\mathbf{x_{0}, y_{0}}$ plane (Fig.~\ref{fig:CFandBtop}), the magnetic field can take 
three inequivalent high symmetry directions: either towards an oxygen that lies \emph{above} the plane ($ 0 \,^{\circ} + n \, 120 \,^{\circ}$), or towards an oxygen that lies \emph{below} the plane ($ 60 \,^{\circ} + n \, 120 \,^{\circ}$), or else precisely in between two oxygens ($ 30 \,^{\circ} + n \, 60 \,^{\circ}$), where in all cases $n$ is an integer. 
The first two directions correspond to the maxima and minima of $m_{z} = g_{J} \mu_{B} \braket{\hat{J}_{z}}$, respectively. The latter direction corresponds to nodes where $m_{z}$ vanishes. 
Curiously, as we noted before, the sign of $m_{z}$ in the first two cases switches between DTO and HTO. We notice that this switching is highly dependent on the precise values of the CF parameters used.

\section{	Finite temperatures 		\label{sec:FiniteTemp}	}

The splitting between ground and first excited state can be very small at low fields (see Fig.~\ref{fig:Splitting}), far smaller than any temperature of experimental interest. 
Since the two states originate (adiabatically) from the splitting of the ground state CF doublet, they will,
in general, preserve the (symmetric) property of being polarised in opposite directions 
(that is, the equivalent behaviour of 
Fig.~\ref{fig:MagnetizAngleTesla}, panels e-f, 
for the first excited state -- not shown -- is simply opposite in sign with respect to that for the ground state). In the absence of a longitudinal field, 
thermal averages between the two will therefore cancel out the single-ion moment.

\subsection{	Magnetic moment	\label{sec:FiniteTempMagnMom}	}

At finite temperature $T$, 
$\braket{\hat{J}_{\alpha}} = \mathrm{Tr}( {\hat{J}_{\alpha} \hat{\rho}} )/\mathrm{Tr}( \hat{\rho} )$, 
where $ \hat{\rho}  = e^{- \hat{\mathcal{H}}/k_{\mathrm{B}} T} $ is the density operator in the microcanonical ensemble, $ \hat{\mathcal{H}}$ is the Hamiltonian in Eq.~\eqref{eq:Ham_Cf_Zeeman}, and 
$ k_{\mathrm{B}} $ is the Boltzmann constant. 

Since $\braket{\hat{J}_{x}}$ and $\braket{\hat{J}_{y}}$ take on negligible values at applied fields below the (trivial) Larmor threshold (see Fig.~\ref{fig:MagnetizAngleTesla}, panels a-d), we focus our discussion on $\braket{\hat{J}_{z}}$. Its behaviour as a function of $\phi$ and $|\mathbf{B}|$ is shown in Fig.~\ref{fig:JzTempAngleTesla}, panels g-h, for $ T = 0.5$~K. 

We find that the anisotropic response survives at intermediate fields, in between a high and a low field threshold. The high field threshold is the (temperature independent) onset of Larmor precession. The low field threshold instead is set by the ground state doublet splitting (Fig.~\ref{fig:Splitting}). The low field threshold is temperature dependent, namely $\sim T^{1/2}$ for HTO and $\sim T^{1/3}$ (i.e., more easily observed) for DTO, according to the results in Sec.~\ref{sec:MagnField}. 

\subsection{	Magnetic susceptibility	\label{sec:FiniteTempSuscept}	}

\begin{figure*}
	\centering
	\subfloat	[HTO \label{fig:SuscNormalCurieHTO}]
			{\includegraphics[trim = 0mm	1mm		4mm 	1mm,	clip,	width=.49\linewidth]{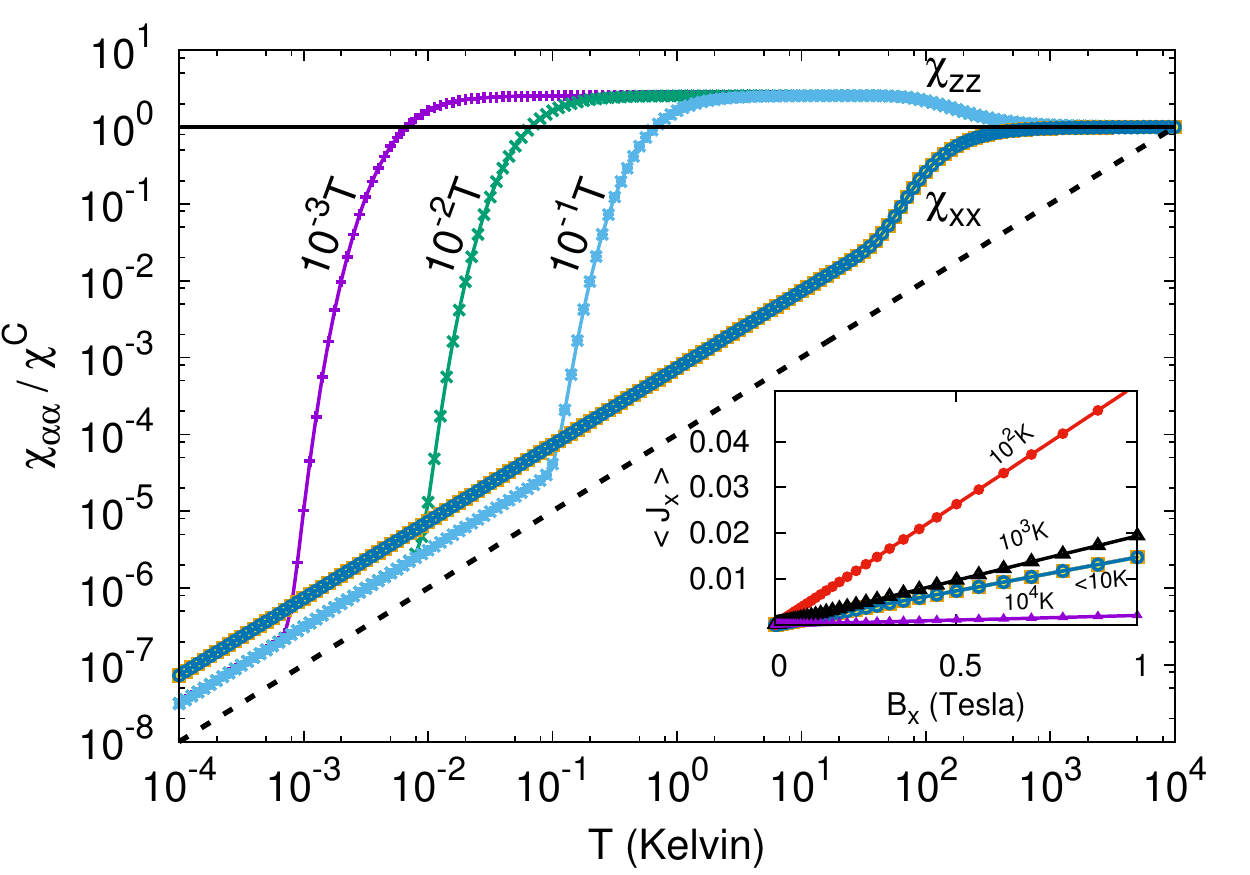}}
	\;
	\subfloat	[DTO \label{fig:SuscNormalCurieDTO}]
			{\includegraphics[trim = 0mm	1mm		4mm 	1mm,	clip,	width=.49\linewidth]{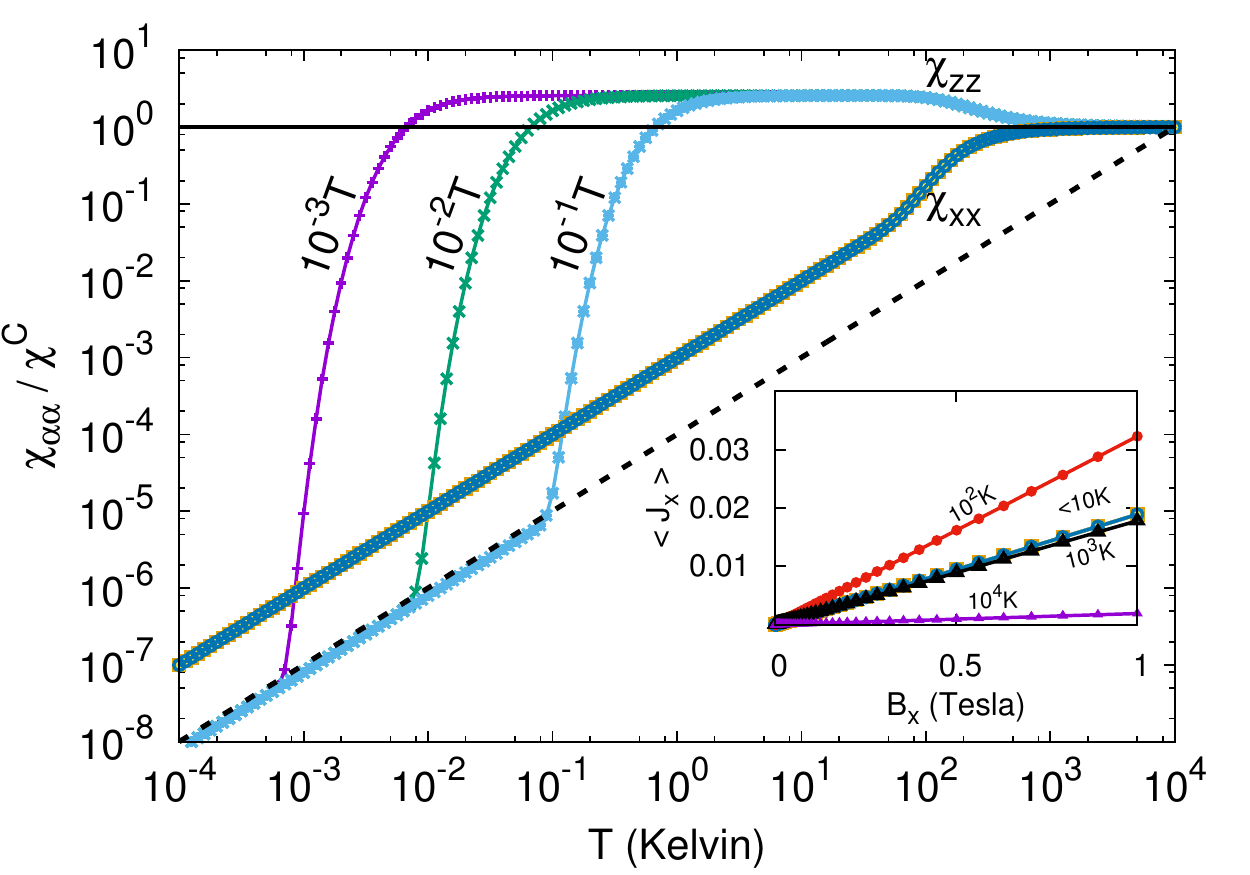}}

	\caption	{	
				\label{fig:SuscNormalCurie}
				Logarithmic plots of $ \chi_{\alpha \alpha}/\chi^{\mathrm{C}} $, with $ \alpha=x,z $, as a function of temperature $ T $ and in presence of a static applied field.  
				When the temperature is lowered the $ xx $-component deviates from the Curie law by decreasing approximately linearly while, 
				the $ zz $-component exhibits an intermediate (higher) plateau 
				($ \chi^{\rm plateau}_{zz}/\chi^{\mathrm{C}} \approx 2.5 $). 
				(Note that $\chi_{\alpha\alpha}/\chi^{\rm C} \propto T$ corresponds, following Eq.~\eqref{eq:SuscRatio}, to $\chi_{\alpha \alpha}$ being T-independent. The dashed line in each panel illustrates a linear behaviour as a guide to the eye.)
	 			Each component is shown for three values of applied fields: $0.001$~T, $0.01$~T, and $0.1$~T. 
				For $ \chi_{xx} $, the three curves overlap almost perfectly, signalling that the susceptibility is field-independent below $0.1$~T. 
				For $ \chi_{zz} $, the three curves overlap only for sufficiently large (field-dependent) threshold temperatures; 
				in the main text we discuss how this behaviour is directly related to the different ground state splittings opened in the crystal-field spectrum by the applied fields. 
				The insets show $ \braket{\hat{J}_{x}} $ vs field $ B_{x} $ at different temperatures, demonstrating a linear regime up to at least $1$~T. 
The temperature-independent susceptibility below $10$~K is reflected in the perfect overlap of the magnetisation curves $ \braket{\hat{J}_{x}} $ vs $  B_{x} $ at these temperatures ($ \chi_{xx} \sim C \, \mu_{0} \,  \mu^{2} / 3 \,  k_{ \mathrm{B} } $, with $C = 0.015 \, \mbox{K}^{-1}$  for HTO and $C = 0.02 \, \mbox{K}^{-1}$ DTO). 
			}
\end{figure*}

The susceptibility of the $ \alpha $-component of the magnetic moment 
with respect to the $\beta$-component of the applied field $B$ is given by 
\begin{equation}
	\begin{split}
		\chi_{\alpha \beta} = \mu_{0} \mu_{\mathrm{B}} g_{J} \frac{\partial \braket{\hat{J}_{\alpha}}}{\partial B_{\beta}} 
		\, .
	\end{split}
\label{eq:SuscGeneral}
\end{equation}
At high temperatures, we expect the system to behave as an ordinary paramagnet,
whose zero-field magnetic susceptibility (per spin) is given by the Curie law 
\begin{equation}
	\begin{split}
		\chi^{\mathrm{C}}_{\alpha \beta}	=	\mu_{0} \frac{\mu^{2}}{ 3 k_{\mathrm{B}} T } \delta_{\alpha \beta} \equiv \chi^{\mathrm{C}},
	\end{split}
\label{eq:SuscCurie}
\end{equation}
where $ \mu^{2} = g_{J}^{2} \mu_{\mathrm{B}}^{2} J(J+1) $. 
Therefore, it is convenient to define the dimensionless quantity 
\begin{equation}
	\begin{split}
		\frac{\chi_{\alpha \beta}}{\chi^{\mathrm{C}}}  =	\frac{ 3 k_{\mathrm{B}} T }{g_{J} \mu_{\mathrm{B}} J(J+1)} 
									\frac{\partial \braket{\hat{J}_{\alpha}}}{\partial B_{\beta}} \,	,	
	\end{split}
\label{eq:SuscRatio}
\end{equation}
whose behaviour is shown for $ \beta=\alpha $ in Fig.~\ref{fig:SuscNormalCurie}. All curves exhibit Curie behaviour at (unphysically) high temperatures. 
(Only $\chi_{xx}$ and $\chi_{zz}$ are shown, as $ \chi_{yy} $ behaves analogously to $\chi_{xx}$.) 

The behaviour of $\chi_{zz}$ at low temperatures is perhaps most remarkable. 
It exhibits a Curie-like intermediate temperature regime where $ \chi^{\rm plateau}_{zz} \approx 2.5 \, \chi^{\mathrm{C}} $, for both HTO and DTO. 
At high temperatures, this regime crosses over to the expected Curie law at an (approximately) field-independent threshold $T \sim 10^2$~K, set by the CF energy gap between the ground state doublet and higher excited states. 
Below this threshold, the system is effectively projected onto its ground state doublet. The magnetic response is thus enhanced since these two states carry the largest magnetic moments of all CF levels. 

In presence of a finite applied magnetic field, as is the case in Fig.~\ref{fig:SuscNormalCurie}, one trivially expects a lower threshold to the effective spin-1/2 Curie behaviour when the temperature becomes smaller than the (linear) Zeeman splitting between the two levels. 
{The system then crosses over to a regime where the
susceptibility is temperature-independent, but finite, corresponding to a
small residual polarizability in the ground state.}

Interestingly, $\chi_{xx}$ displays a similar behaviour, in spite of the fact that the splitting between the two lowest-lying states in a transverse field is now much smaller than temperature. In this case the temperature-independent regime extends all the way to $T \sim 10$~K.

\section{	Summary and discussion		\label{sec:Discussion}	}

We have presented a detailed study of the single-ion behaviour in spin ice HTO and DTO in presence of an applied magnetic field, based on the full description of the single-ion crystal-field Hamiltonian. We have considered both zero and finite temperature, and focused in particular on the case of a field transverse to the local easy axis. 
We find that the Kramers vs non-Kramers nature of HTO vs DTO results in a different perturbative behaviour at small fields. 

We also present a detailed study of the static magnetic response to a transverse field, which highlights the anisotropy of the system. We find that as a function of the in-plane direction of the field, off-diagonal components of the $g$-tensor become non-vanishing (namely, a purely transverse field in the $xy$-plane induces a longitudinal response along the local $z$ axis).  

Within the classical exchange and dipolar Hamiltonian approximation, the action of all other ions on a given one is an effective magnetic field, whose strength and direction were studied in Ref.~\onlinecite{Sala:2012}. 
The transverse component of this effective field can be thought of as a potential source of quantum dynamics in spin systems, and the corresponding ground state splitting studied in the present paper corresponds in this view to an inverse characteristic time scale. 
It is then remarkable to notice that -- in spite of these simplifying assumptions -- the resulting time scales for HTO and DTO are consistent with the ones observed in experiments~\cite{Snyder:2004}.  

There are a number of natural directions for future work. 
One is towards a yet more microscopic picture, 
going beyond the effective field approximation we have employed here, 
by determining the actual exchange Hamiltonian for example via a superexchange calculation. 
Another lies in considering the interplay of the spin degrees of freedom in the same spirit as we have considered the coupling to an external field. 
For the case of magnetoelastic couplings, a simple calculation in this spirit was reported in Ref.~\onlinecite{Erfanifam:2014}.

Indeed,	the issue of coupling to non-magnetic degrees of freedom is of
relevance given the remarkably long millisecond-timescale of the spin
flip and the small splitting of the ground doublet. These are below
$10^{-5}K$ in temperature units, well below the scale at which
experiments are conducted. Understanding the spin tunnelling process
in the presence of a coupling to the `hot' environment is therefore an
interesting exercise, close in spirit to the study of molecular
magnets, which may be of relevance to some of the unexplained features
of the	slow low-temperature dynamics of spin ice.
 
Many of our results can be tested experimentally, perhaps best in heavily Y-diluted systems in an externally applied magnetic field, to reduce the added complexity of spin-spin interactions. 
Fig.~\ref{fig:JzTempAngleTesla} shows, that the anisotropic response of a RE ion in spin ice could be observed at temperatures $\sim 100$~mK under externally-applied fields $\sim 10$~T. 
{At lower fields, the induced magnetic moment is much lower, but, as the insets of Fig. 8 show, it could still be detectable, for example by muon spin rotation.}
Such experiments could provide a quantitative validation of the present description, which is a crucial step towards gaining further insight into the quantum dynamics of spin ice materials.

\section{	Acknowledgments		\label{sec:Acknowledgments}	}

This work was supported in part by EPSRC Grant No.~EP/K028960/1 (C.C.) and in part by the Helmholtz Virtual Institute ``New States of Matter and Their Excitations'' and by the EPSRC NetworkPlus on ``Emergence and Physics far from Equilibrium''.
B.T.  was supported by a PhD studentship from HEFCE and SEPnet.  J.Q. was supported by a fellowship from STFC and SEPnet.
Statement of compliance with EPSRC policy framework on research data: this publication reports theoretical work that does not require supporting research data.

The authors thank S.~Giblin, J.~Chalker, T.~Fennel, S.T.~Bramwell, C.~Hooley, and P.~Strange for useful discussions. 
Very special thanks go to G.L.~Pascut for the support given to B.T. on the study of crystal-field theories.
B.T. is also very grateful for the hospitality during his stay at the Max Planck Institute 
for the Physics of Complex Systems in Dresden, where part of the work was carried out.

\appendix

\section	{Oxygen environment 	\label{sec:App_MagnPyro}		}

The crystal-field interactions, i.e., the Stark effect due to the negative charges of the oxygen ions, deeply affect the single-ion quantum states. 
The physics dictated by the crystalline fields of the oxygens is so fundamental that, in the context of magnetic pyrochlore oxides, 
often it is preferable to use the expression A$_{2}$B$_{2}$O$_{6}$O$'$, instead of A$_{2}$B$_{2}$O$_{7}$, 
simply to emphasise the role played by the oxygens according to their crystallographic and ligand character.
Referring to a given RE$^{3+}$ (A) site, e.g., a Ho$^{3+}$ ion, the oxygens are arranged 
around it in an anti-prismatic fashion which is often referred to as a distorted cube (see Fig.~\ref{fig:CFenvironment} in the main text).

The level of distortion is, however, huge compared to an ideal cube,
as the two O1 oxygens form a linear O-A-O stick 
oriented normal to the average plane of the remaining six O2 oxygens arranged in triangles above and below the central A ion. 
The A-O1 and A-O2 bond distances are different: the former, $\sim 2.2$~{\AA}, 
is amongst 
the shortest bonds ever found in nature; the latter can vary depending on the compound, 
although in general it is between $2.4$ and $2.5$~{\AA}~\cite{Subramanian:1993,Gardner:2010}. 
This implies that each RE$^{3+}$ ion is characterised by a very pronounced axial symmetry along the local $\langle111\rangle$ axis 
which joins the two centres (O1 sites) of the tetrahedra, through the magnetic ion sitting at the shared vertex 
(see Fig.~\ref{fig:CFenvironment(a)}).
The axial symmetry is affected by the anti-prismatic arrangement of the O2 ions with respect to the central RE$^{3+}$ ion.
These, as shown in Fig.~\ref{fig:CFenvironment} in the main text,	
are grouped in triangles lying on planes, above and below the RE$^{3+}$ ion, which are parallel to each other. 

\section	{Stevens operators 	\label{sec:App_MagnPyro_SOps}		}

The Stevens operators $\hat{O}_{q}^{k}$ in Eq.~\eqref{eq:HamCFpyroStevensShort},
expressed in terms of the angular momentum operators, can be written as: 
\begin{equation}
	\begin{split}
	\hat{O}_0^2	=	\,	& 	3   \hat{J}^2_z   -      \hat{\mathbf{J}}^2										\\[0.2cm]
	\hat{O}_0^4	=	\,	&	35 \hat{J}^4_z	+ 25\hat{J}^2_z		- 30\hat{\mathbf{J}}^2\hat{J}^2_z 
								- 6\hat{\mathbf{J}}^2		+ 3 \hat{\mathbf{J}}^4 							\\[0.2cm]
	\hat{O}_3^4	=	\,	& 	\frac{1}{4} \left\{		(\hat{J}_{+}^{3}+\hat{J}_{-}^{3})		,		\hat{J}_z	\right\} 	\\[0.2cm]
	\hat{O}_0^6	=	\,	& 	231 \hat{J}^6_z		+ \left(	
																			735
																			- 315\hat{\mathbf{J}}^2
																   \right)	
																   \hat{J}^4_z 						\\[0.1cm]
						&										\quad	+ \left(	
																			  294	
																			- 525\hat{\mathbf{J}}^2
																			+105 \hat{\mathbf{J}}^4
																		  \right)
																\hat{J}^2_z						\\[0.1cm]
						&	\qquad 	-  60 \hat{\mathbf{J}}^2		
									+ 40 \hat{\mathbf{J}}^4	
																-5 \hat{\mathbf{J}}^6					\\[0.2cm]
	\hat{O}_3^6	=	\,	&	\frac{1}{4} 		\left\{ 
																\left(\hat{J}_{+}^{3}+\hat{J}_{-}^{3} \right), 
																\left(	11 \hat{J}_z^3 
																		+ 59 \hat{J}_z 		
																		- 3 \hat{\mathbf{J}}^2 \hat{J}_z 
																\right) 
														\right\}; 									\\[0.2cm]
	\hat{O}_6^6	=	\,	&	\frac{1}{2} \left( \hat{J}_+^6 + \hat{J}_-^6 \right) \, ,
	\end{split}
	\label{eq:StevensOper}
\end{equation}
where $\hat{J}_{\pm} = \hat{J}_x \pm i\hat{J}_y$ and the anticommutator 
$\left\{ \hat{A}, \hat{B} \right\}= \hat{A} \hat{B} + \hat{B} \hat{A}$.

Since each $\hat{O}_{q}^{k}$ is a function of $\hat{\mathbf{J}} = \left( \hat{J}_{x},\hat{J}_{y},\hat{J}_{z} \right)$, 
the total angular momentum operator of the magnetic ion,
the single-ion CF states can be conveniently expressed in terms of $ \ket{J,M_{J}} $, 
where $ J,M_{J} $ are the quantum numbers for, respectively, the total angular momentum and its projection along the local $\langle111\rangle$ axis. 
A list of the matrix elements of the Stevens operators in the $ \ket{J,M_{J}}$ basis is given in Ref.~\onlinecite{Hutchings:1964}.

\section	{	Derivation of the Stevens crystal-field parameters 	\label{sec:App_MagnPyro_CF}		}

The interaction between a magnetic RE$^{3+}$ ion and its surrounding crystalline environment
is usually described starting from the simple Hamiltonian 
\begin{equation}
	\hat{\mathcal{H}}_{CF}	 =	- \sum_{i} |e_i|	V_{CF} (	\hat{\mathbf{r}}_{i}	)	\, ,
\label{eq:HamCFgeneral}
\end{equation}
where $ V_{CF}$ represents the crystal-field potential from to the surrounding ions
acting on the electrons in the unfilled shells of the central RE.
Each $i$-th electron feels a potential $ V_{CF}({\mathbf{r}_{i}}) \equiv V_{CF}(r_i, \theta_i, \phi_i) $ at position $ \boldsymbol{r}_i $.
To study the crystal-field interaction it is convenient to make use of spherical coordinates centred on the RE site
because of the spherical symmetries of the electrons of an atomic system~\cite{Hutchings:1964}. It is customary to write the Hamiltonian in terms of tensor operators. The tensor operator for the $i$-th electron is
\begin{equation}
\hat{C}_q^k (i) = \sqrt{\frac{4\pi}{2k+1}} \; \hat{Y}_k^q	(\theta_{i}, \phi_{i})	\, , 
\end{equation} 
and obeys the same transformation rules as the spherical harmonics.
In terms of these, 
the CF Hamiltonian for a magnetic ion in a crystalline $D_{3d}$ point-group symmetry reads~\cite{Hufner:1978} 
\begin{equation}
	\begin{split}
	\hat{\mathcal{H}}_{\mathrm{CF}} = \,		&		B_0^2 	\hat{C}_0^2	+	B_0^4 	\hat{C}_0^4	+	B_3^4 ( \hat{C}_{3}^4	-	\hat{C}_{-3}^4)	\\[0.1cm]
					      				&	+	B_0^6 	\hat{C}_0^6	+	B_3^6 (	\hat{C}_3^6	-	\hat{C}_{-3}^6) 
																+	B_6^6 (	\hat{C}_6^6	+	\hat{C}_{-6}^6) \, ,
	\end{split}
	\label{eq:HamCF_rosenkranz}
\end{equation}
Here the sum over the 4-$f$ electrons ($ \sum_{i=1}^{n}$) is omitted together with the index $i$ for simplicity. 

The $B_q^k$ parameters encapsulate the effect of the surrounding charges. Eq.~(\ref{eq:HamCF_rosenkranz}) is thus an alternative notation to the one based on Stevens operators, Eq.~(\ref{eq:HamCFpyroStevensShort}). 
The latter is convenient for RE$^{3+}$ ions, where $\ket{J, M_{J}}$ is a good basis for the quantum states of the correlated 4-$f$ electrons, because they are explicit functions of the 
angular momentum operators $ \hat{\mathbf{J}}^2, \hat{J}_{z}, \hat{J}_{+},\hat{J}_{-} $~\cite{Hufner:1978}. 
The $\widetilde{B}_q^k$ are related to the $B_q^k$ by means of the following expressions:
\begin{equation}
\begin{split}
	\widetilde{B}_0^k   = & \sqrt{\frac{4 \pi }{2k+1}} 		\theta_k D_0^k B_0^k 	\\[0.1cm]
	\widetilde{B}_q^k   = & (-1)^q \sqrt{\frac{8 \pi }{2k+1}}\theta_k D_q^k B_q^k 	\quad , \quad 	\text{for} \ q>0 \, .
\end{split}
\label{eq:CFparamBtoGam}
\end{equation}
The $ D_q^k $ are the factors outside the square brackets $[\dots]$ in the
list of tesseral harmonics in Cartesian coordinates in Table IV of Ref.~[\onlinecite{Hutchings:1964}]. 
The $ \theta_{k}$ (with $k=2,4,6$; $ \theta_2=\alpha_J, \theta_4=\beta_J, \theta_6=\gamma_J $)
calculated by Stevens for different RE ions~\cite{Stevens:1952} are given in Table VI of the same Ref.~[\onlinecite{Hutchings:1964}].
In Table \ref{tab:thetas} we reproduce the values for $\alpha_J,\beta_J,\gamma_J$
for the two magnetic ions Ho$^{3+}$ and Dy$^{3+}$ in spin ice materials.

\begin{table}[ht]	\centering	
\setlength{\tabcolsep}{0.5cm}
\def\arraystretch{2.0}%
\begin{tabular}{ c | c | c |}
							&	Ho$^{3+}$ 						&	Dy$^{3+}$						\\[0.2cm]	\hline
	$\displaystyle{\alpha_J}$ 		&	$\displaystyle{\frac{-1}{450}}$ 			& 	$\displaystyle{\frac{-2}{315}}$		\\[0.2cm]	\hline
	$\displaystyle{\beta_J}$ 		&	$\displaystyle{\frac{-1}{30030}}$ 		& 	$\displaystyle{\frac{-8}{135135}}$	\\[0.2cm]	\hline
	$\displaystyle{\gamma_J}$	&	$\displaystyle{\frac{-5}{3864861}}$		& 	$\displaystyle{\frac{4}{3864861}}$	\\[0.2cm]	\hline	
\end{tabular}
	\caption	{	The $ \theta_k $ values 
				(respectively $\alpha_J$, $\beta_J$, $\gamma_J$ for $k=2,4,6$) 
				for  Holmium and Dysprosium trivalent ions~\cite{Hutchings:1964}.}
	\label{tab:thetas}
\end{table}						

Experimental techniques based on inelastic neutron scattering are the most suitable to measure accurately the crystal-field energies in real compounds. From these measurements a reliable estimate of the CF parameters can be inferred beyond the level of accuracy allowed by the point-charge approximation~\cite{Hufner:1978,Takegahara:2000,Mulak:2000}. 

The crystal-field energies and parameters common in the literature of spin ice materials
are based mainly on the experiment presented by Rosenkranz et al.
in Ref.~\onlinecite{Rosenkranz:2000}.
There, the neutron scattering measurement of all the CF energy levels allowed a complete parametrisation of the Hamiltonian in Eq.~\eqref{eq:HamCF_rosenkranz}.
The full list of the $B_{k}^{q}$ parameters for HTO found in that reference is reproduced in the first column of Table~\ref{tab:spiniceCFparamTensor}. 
Similarly for DTO, the second column of Table~\ref{tab:spiniceCFparamTensor} gives the $B_{k}^{q}$ suggested in Ref.~\onlinecite{Malkin:2010}
as an interpolation of the values known for Ho$_{2}$Ti$_{2}$O$_{7}$ and Tb$_{2}$Ti$_{2}$O$_{7}$. 
However, to the best of our knowledge no neutron scattering experiments have been carried out successfully to determine the CF parameters of DTO.
The corresponding $\widetilde{B}_{k}^{q}$ parameters, listed in Table~\ref{tab:spiniceCFparamStev} for the Hamiltonian Eq.~\eqref{eq:HamCFpyroStevensShort} in the main text,
follow from Eq.~\eqref{eq:CFparamBtoGam} and Table~\ref{tab:spiniceCFparamTensor}.

\begin{table}[ht]	\centering
\setlength{\tabcolsep}{0.3cm}
\def\arraystretch{1.5}%
\begin{tabular}{ c | c c |}
					&		HTO	(meV)	&		DTO	(meV)	\\[0.1cm]	\hline
		$B_0^2$ 		&		68.2			&		51.1			\\[0.1cm]	\hline
		$B_0^4$		&		274.8		&		306.2		\\[0.1cm]	\hline	
		$B_3^4$		&		83.7			&		90.5			\\[0.1cm]	\hline	
		$B_0^6$ 		&		86.8			&		100.4		\\[0.1cm]	\hline
		$B_3^6$ 		&		-62.5			&		-74.4			\\[0.1cm]	\hline
		$B_6^6$		&		101.6		&		102.9		\\
		\hline
\end{tabular}
	\caption
				{	Crystal-field parameters $B_{k}^{q}$ for the tensor operators formalism.
					The parameters for HTO have been measured by means of inelastic neutron scattering in Ref.~\onlinecite{Rosenkranz:2000}. 
					The ones for DTO were derived as an interpolation of the parameters   
					known for Ho$_{2}$Ti$_{2}$O$_{7}$ and Tb$_{2}$Ti$_{2}$O$_{7}$ in Ref.~\onlinecite{Malkin:2010}.
				}
\label{tab:spiniceCFparamTensor}						
\end{table}						

\section{	Degenerate Perturbation Theory		\label{sec:App_MagnPyro_PerturTheory}	}

The results in Sec.~\ref{sec:MagnField_ExactDiag}, 
namely Eqs.~(\ref{eq:power_law_Ho3+}-\ref{eq:Delta01Perturb}), 
follow from conventional degenerate perturbation theory (see e.g., Ref.~\onlinecite{Bransden:2000}). 
In this Appendix we outline the main steps of the derivation for convenience. 
This will help understand the role that the symmetries of the unperturbed CF Hamiltonian play in determining the perturbative behaviour, as discussed in the main text and in detail in App.~\ref{sec:App_MatElemPerturb}. 
(The interested reader can find all the details of the calculations in  Ref.~\onlinecite{TomaselloPhDThesis}.) 

We use the notation 
\begin{equation}
\hat{H}=\hat{H}_{0}+\lambda\hat{V} 
\, , 
\label{eqA:H_full}
\end{equation}
where $\lambda$ is a small (real) parameter tuning the strength of the perturbation $\hat{V}$, and where the eigenstates and eigenvalues of $\hat{H}_{0}$ are 
\begin{equation}
\hat{H}_{0}\ket{\psi_{n}^{(0)}}=E_{n}^{(0)}\ket{\psi_{n}^{(0)}}
\, . 
\label{eqA:Schr_0}
\end{equation}
Here $\hat{H_{0}}\equiv\hat{\mathcal{H}}_{\mathrm{CF}}$, in Eq.(\ref{eq:HamCFpyroStevensShort}) of the main text, and 
$\hat{V}$ is the applied magnetic field, see Eq.~\eqref{eq:perturbation} and Eq.~\eqref{eq:HamPerturb}. 
We focus on the case of interest where the first two energy levels are exactly degenerate ($E_{1}^{(0)} = E_{0}^{(0)}$). Of course, in this case the choice of basis for the ground state doublet, $\ket{\psi_{0}^{(0)}}$ and $\ket{\psi_{1}^{(0)}}$, is not unique. 

Expanding both eigenstates and eigenvalues of the perturbed Hamiltonian in powers of the parameter $\lambda$, one obtains the contributions to the GS doublet splitting order by order. The form of the contribution at a given order depends of course on whether the two levels did or did not split at lower order.

For notational convenience, it is useful to define the matrix elements $V_{n,m}=\bra{\psi_{n}^{(0)}}\hat{V}\ket{\psi_{m}^{(0)}}$.

\subsection{	First  order	\label{sec:App_MagnPyro_PertTh_1ord}	}

The first order contribution is bookwork, 
\begin{equation}
\sqrt{(V_{0,0}-V_{1,1})^{2}+4|V_{0,1}|^{2}} 
. 
\end{equation}
However, (main text and App.~\ref{sec:App_MatElemPerturb}) both DTO and HTO have $V_{0,1}=V_{0,0}=V_{1,1}=0$ so that the degeneracy is not resolved at first order in $\lambda$.

\subsection{	Second order	\label{sec:App_MagnPyro_PertTh_2ord}	}

We focus on the case of interest where $V_{0,0}=V_{1,1}=V_{1,0}=V_{0,1}=0$. 
After a few lines of algebra one obtains that the second order (GS doublet) contribution to the splitting takes the form 
\begin{equation}
\sqrt{\Bigg(\sum_{k>1}\frac{|V_{0,k}|^{2}-|V_{1,k}|^{2}}{\Delta E_{k}^{(0)}}\Bigg)^{2}+4\Bigg|\sum_{k>1}\frac{V_{0,k}V_{k,1}}{\Delta E_{k}^{(0)}}\Bigg|^{2}} 
, 
\label{eqA:E2pm}
\end{equation}
where $\Delta E_{k}^{(0)}$ is the energy difference between the GS doublet and the $k^{\rm th}$ excited state of the unperturbed Hamiltonian $\hat{H}_0$. 
The summation over $k > 1$ spans all unperturbed states other than the GS doublet. 

As discussed in the main text and in App.~\ref{sec:App_MatElemPerturb}, we find that all excited states that form doublets in the unperturbed spectrum amount to a vanishing contribution to the second order GS doublet splitting in Eq.~\eqref{eqA:E2pm}. 
This is not the case for singlets, which are present in the spectrum of non-Kramers HTO, where splitting occurs at second order and Eq.~\eqref{eqA:E2pm} is in good agreement with the numerical solution at sufficiently small values of the perturbation parameter $\lambda$ (see main text). 
On the other hand, Kramers theorem forbids the appearance of singlets in DTO resulting  in a vanishing second order splitting.

\section{	Matrix elements of the perturbation \label{sec:App_MatElemPerturb} }

In this Appendix we give details of the calculation of the matrix elements 
$\widetilde{V}_{n,m}=\bra{\psi_{n}^{(0)}}\hat{\widetilde{V}}\ket{\psi_{m}^{(0)}}$, 
where 
$\ket{\psi_{m}^{(0)}}$ are the eigenstates of the unperturbed CF Hamiltonian, 
and the dimensionless operator 
\begin{equation}	\label{eq:Vtilde}
	\begin{split}
		\hat{\widetilde{V}}=	e^{ - i \phi } \hat{ J}_{+} +
						e^{ + i \phi} \hat{ J}_{-}
	\end{split}
\end{equation}
represents the applied magnetic field (perturbation) purely transverse 
to the local quantisation axis of the {RE}-ion (see  Eq.~\eqref{eq:perturbation} in the main text). 
Namely, the operator 
$\hat{V}\equiv\mathbf{\mathfrak{\mathcal{E}}_{\mathrm{CF}}} \, \hat{\mathbf{J}}\cdot\mathbf{B}/\mathbf{|B|}$
in Sec.~\ref{sec:MagnField_ExactDiag} relates to Eq.~\eqref{eq:Vtilde} via 
$\hat{\widetilde{V}}=2\hat{V} / \mathfrak{\mathcal{E}}_{\mathrm{CF}}$. 
The perturbative regime corresponds to field values within the initial power-law behaviour of the splitting observed in Fig.~\ref{fig:Splitting}. 

The quantum states for HTO and DTO are represented in Table~\ref{tab:HTOstates} and Table~\ref{tab:DTOstates}, respectively. 
\begin{table*}[]\centering
\setlength{\tabcolsep}{0.2cm}
\def\arraystretch{1.5}%
\begin{tabular}{ l || c c | c c | c  c |  c | c |}
HTO			& $ \ket{\psi_{A}^{(0)}} $ & $ \ket{\psi_{A+1}^{(0)}} $ 
												   & $ \hat{\widetilde{V}}\ket{\psi_{A}^{(0)}} $ & $ \hat{\widetilde{V}}\ket{\psi_{A+1}^{(0)}} $ 
												   																   & $ \ket{\psi_{B}^{(0)}} $ & $ \ket{\psi_{B+1}^{(0)}} $ & $ \ket{\psi_{s}^{(0)}} $  & $ \ket{\psi_{s'}^{(0)}} $ \\[0.1cm]
\hline
$\ket{8}$  		& $ a_{8}	$ 		     & $ 0 $ 			   & $ 0 $ 							   & $ -j_{7}^{+} \, a_{-7}$ 				   & $ b_{8} $ 			   & $ b_{-8} $			  & $ 0 $ 				& $ 0 $ 			      \\[0.1cm]
$\ket{7}$  		& $  0	$ 		     & $ -a_{-7}$ 			   & $ j_{8}^{-} \, a_{8}  $				   & $ 0 $							   & $ b_{7} $ 			   & $ b_{-7} $			  & $ 0 $ 				& $ 0	 $ 			      \\[0.1cm]
$\ket{6}$  		& $  0	$		     &	$ 0 $				   & $ j_{5}^{+} \, a_{5} $				   & $ -j_{7}^{-} \, a_{-7} $				   &	 $  0	$			   & $ 0 $				  & $ s_{6} $			& $ s'_{6} $		      \\[0.1cm]
$\ket{5}$		& $  a_{5} $ 		     & $ 0 $				   & $ 0 $ 							   & $  j_{4}^{+} \, a_{-4} $ 				   & $ b_{5} $			   & $	 -b_{-5} $			  & $	 0 $				& $ 0	 $			      \\[0.1cm]
$\ket{4}$		& $  0 $			     & $ a_{-4} $			   & $	 j_{5}^{-} \, a_{5}  $				   & $	 0 $							   & $ b_{4} $			   & $	 -b_{-4} $			  & $	 0 $				& $ 0	 $			      \\[0.1cm]
$\ket{3}$		& $  0 $			     & $ 0 $				   & $ j_{2}^{+} \, a_{2} $				   & $ j_{4}^{-} \, a_{-4} $				   & $  0	 $			   & $  0 $				  & $	 s_{3}$			& $ s'_{3}	$		      \\[0.1cm]
$\ket{2}$		& $  a_{2} $		     & $ 0 $				   & $ 0 $							   & $-j_{1}^{+} \, a_{-1}$				   & $ b_{2} $			   & $	 b_{-2} $			  & $	 0 $				& $ 0 $			      \\[0.1cm]
$\ket{1}$		& $  0 $			     & $ -a_{-1} $			   & $	 j_{2}^{-} \, a_{2}  $				   & $ 0 $							   & $ b_{1} $			   & $ b_{-1} $			  & $	 0 $				& $ 0	 $			      \\[0.1cm]
$\ket{0}$		& $  0 $			     & $ 0 $ 			   & $	 j_{-1}^{+} \, a_{-1}$				   & $ -j_{1}^{-} \, a_{-1} $				   & $ 0 $				   & $ 0 $				  & $	 s_{0} $			& $ 0	 $			      \\[0.1cm]
$\ket{-1}$		& $  a_{-1}$		     & $ 0 $				   & $	 0 $							   & $	 j_{-2}^{+} \, a_{2}$				   & $ b_{-1}$			   & $ -b_{1} $			  & $ 0 $				& $ 0 $			      \\[0.1cm]
$\ket{-2}$		& $ 0	 $			     & $ a_{2} $			   & $ j_{-1}^{-} \, a_{-1} $				   & $ 0 $							   &	 $ b_{-2}$			   & $ -b_{2} $			  & $ 0 $				& $ 0	 $			      \\[0.1cm]
$\ket{-3}$		& $ 0 $			     & $ 0 $				   & $ j_{-4}^{+} \, a_{-4}$				   & $ j_{-2}^{-} \, a_{2} $				   & $ 0 $				   & $	  0 $				  & $ -s_{3} $			& $ s'_{3} $		      \\[0.1cm]
$\ket{-4}$		& $ a_{-4} $		     & $ 0 $				   & $ 0 $							   & $	-j_{-5}^{+} \, a_{5}$				   & $ b_{-4} $			   & $ b_{4} $			  & $ 0 $				& $ 0	 $			      \\[0.1cm]
$\ket{-5}$		& $ 0 $			     & $ -a_{5} $			   & $ j_{-4}^{-} \, a_{-4} $				   & $ 0 $							   & $ b_{-5} $			   & $ b_{5} $			  & $ 0 $				& $ 0	 $			      \\[0.1cm]
$\ket{-6}$		& $ 0 $			     & $ 0 $				   & $ j_{-7}^{+} \, a_{-7} $				   & $ -j_{-5}^{-} \, a_{5} $ 				   & $ 0 $				   & $ 0 $				  & $ s_{6} $			& $ -s'_{6} $		      \\[0.1cm]
$\ket{-7}$		& $ a_{-7} $		     & $ 0 $				   & $ 0 $							   & $ j_{-8}^{+} \, a_{8} $				   & $ b_{-7} $			   & $ -b_{7} $			  & $ 0 $				& $ 0	 $			      \\[0.1cm]
$\ket{-8}$		& $ 0	 $			     & $ a_{8} $			   & $ j_{-7}^{-} \, a_{-7} $				   & $ 0 $							   & $ b_{-8} $			   & $ -b_{8} $			  & $ 0 $				& $ 0	 $			      \\[0.1cm]
\hline
\end{tabular}
	\caption	{	The coefficients for the decomposition of the single-ion states of HTO
				with respect to the angular momentum eigenstates $\ket{M_{J}}$.
				A generic state $\ket{\psi}$ at the top of a column of coefficients $C_{M_{J}}$ 
				is given by a superposition $ \ket{\psi} = \sum_{M_{J}} C_{M_{J}} \ket{M_{J}}$.
				It is useful to distinguish between $A$-type and $B$-type doublet states, as well as singlets $s$ and $s'$. 
				The ground state doublet $ \ket{\psi_{0}^{(0)}},\, \ket{\psi_{1}^{(0)}} $ is of the $A$-type; 
				the coefficients are $\{ a_{8}=0.982, a_{5}=0.156, a_{2}=0.065, a_{-1}=0.071, a_{-4}=0.049, a_{-7}=0.006 \}$
				and account for the strong Ising anisotropy characteristic of the spin ice single-ion physics. 
				The second and third column from the left are
				the first order terms obtained by applying $\hat{\widetilde{V}}$ to an $A$-type doublet. 
			}
\label{tab:HTOstates}
\end{table*}						
\begin{table*}[]\centering
\setlength{\tabcolsep}{0.2cm}	
\def\arraystretch{1.5}%
\begin{tabular}{ l || c c | c c |  c c |}
DTO			& $ \ket{\psi_{A}^{(0)}} $ 	& $ \ket{\psi_{A+1}^{(0)}} $ & $ \hat{\widetilde{V}}\ket{\psi_{A}^{(0)}} $	 & $ \hat{\widetilde{V}}\ket{\psi_{A+1}^{(0)}} $	& $ \ket{\psi_{B}^{(0)}} $	& $ \ket{\psi_{B+1}^{(0)}} $	\\[0.1cm]
\hline
$\ket{15/2}$  	& $ a_{15/2}	$ 		& $ a_{-15/2}  $			  & $ 0 $								 & $ 0 $	 							 & $ 0 $ 				& $ 0	 $				\\[0.1cm]
$\ket{13/2}$  	& $ 0 $ 				& $ 0	 $ 				  & $ j^{-}_{15/2} \, a_{15/2} $				 & $ j^{-}_{15/2} \, a_{-15/2} $				 & $ b_{13/2}	$ 		& $ b_{-13/2}  $			\\[0.1cm]
$\ket{11/2}$  	& $ 0 $ 				& $ 0	 $				  & $ j^{+}_{9/2}  \, a_{9/2}$ 				 & $ - j^{+}_{9/2} \,a_{-9/2} $				 & $ b_{11/2}	$ 		& $ -b_{-11/2}  $		\\[0.1cm]
$\ket{9/2}$	& $ a_{9/2}	$ 		& $ -a_{-9/2}   $			  & $ 0 $ 								 & $ 0 $								 & $ 0 $ 				& $ 0	 $				\\[0.1cm]
$\ket{7/2}$	& $ 0 $				& $ 0	 $				  & $ j^{-}_{9/2}  \,  a_{9/2}	$ 				 & $ -j^{-}_{9/2} \, a_{-9/2} $				 & $ b_{7/2}	$ 		& $ -b_{-7/2}  $			\\[0.1cm]
$\ket{5/2}$	& $ 0 $ 				& $ 0	 $				  & $ j^{+}_{3/2} \,  a_{3/2}$ 				 & $ j^{+}_{3/2} \,  a_{-3/2}$				 & $ b_{5/2}	$ 		& $ b_{-5/2}  $			\\[0.1cm]
$\ket{3/2}$	& $ a_{3/2}	$		& $ a_{-3/2}   $			  & $ 0 $ 								 & $ 0	 $							 & $ 0 $ 				& $ 0	 $				\\[0.1cm]
$\ket{1/2}$	& $ 0 $				& $ 0	 $				  & $ j^{-}_{3/2}  \, a_{3/2}$ 				 & $ j^{-}_{3/2} \,  a_{-3/2}$				 & $ b_{1/2}	$ 		& $ b_{-1/2}  $			\\[0.1cm]
$\ket{-1/2}$	& $ 0 $ 				& $ 0	 $				  & $ j^{+}_{-3/2}\, a_{-3/2}$ 				 & $ j^{+}_{-3/2}  \, -a_{3/2} $				 & $ b_{-1/2}	$ 		& $ -b_{1/2}  $			\\[0.1cm]
$\ket{-3/2}$	& $ a_{-3/2}	$		& $ -a_{3/2}    $			  & $ 0 $ 								 & $ 0	 $							 & $ 0 $ 				& $ 0	 $				\\[0.1cm]
$\ket{-5/2}$	& $ 0 $				& $ 0	 $				  & $ j^{-}_{-3/2} \, a_{-3/2}$ 				 & $ -j^{-}_{-3/2} \,  a_{3/2}$				 & $ b_{-5/2}	$ 		& $ -b_{5/2}  $			\\[0.1cm]
$\ket{-7/2}$	& $ 0 $ 				& $ 0	 $				  & $ j^{+}_{-9/2}\, a_{-9/2}$ 				 & $ j^{+}_{-9/2} \, a_{9/2}$				 & $ b_{-7/2}	$ 		& $ b_{7/2}  $			\\[0.1cm]
$\ket{-9/2}$	& $ a_{-9/2}	$		& $ a_{9/2}     $			  & $ 0 $ 								 & $ 0	 $							 & $ 0 $ 				& $ 0	 $				\\[0.1cm]
$\ket{-11/2}$	& $ 0 $				& $ 0	 $				  & $ j^{-}_{-9/2} \, a_{-9/2} $ 				 & $ j^{-}_{-9/2} \, a_{9/2}$					 & $ b_{-11/2}	$ 		& $ b_{11/2}  $			\\[0.1cm]
$\ket{-13/2}$	& $ 0 $ 				& $ 0	 $				  & $ j^{+}_{-15/2}\,a_{-15/2}  $ 				 & $ -j^{+}_{-15/2} \, a_{15/2}$				 & $ b_{-13/2}	$ 		& $ -b_{13/2}  $			\\[0.1cm]
$\ket{-15/2}$	& $ a_{-15/2}	$		& $ -a_{15/2}  $			  & $ 0 $ 								 & $ 0	 $							 & $ 0 $ 				& $ 0	 $				\\[0.1cm]
\hline
\end{tabular}
	\caption	{	The coefficients for the decomposition of the single-ion states of DTO
				with respect to the angular momentum eigenstates $\ket{M_{J}}$.
				The CF states for DTO are all doublets. 
				These are either of type $A$: $ \ket{\psi_{A}^{(0)}} $ and $ \ket{\psi_{A+1}^{(0)}}$ in the first and second column, respectively; 
				or of type $B$: $ \ket{\psi_{B}^{(0)}} $ and $ \ket{\psi_{B+1}^{(0)}}$  in the fifth and sixth column, respectively. 
				The ground state doublet $ \ket{\psi_{0}^{(0)}}, \ket{\psi_{1}^{(0)}} $ belongs to the $A$-type states;
				the coefficients are
				$\{ a_{15/2}=0.983, a_{9/2}=-0.171, a_{3/2}=0.044, a_{-3/2}=0.044, a_{-9/2}=0.008, a_{-15/2}=0 \}$.
				The second and third column are
				the first order terms obtained by applying $\hat{\widetilde{V}}$ to a pair of $A$-doublet states. 
			}
\label{tab:DTOstates}
\end{table*}						
Each state $\ket{\psi}$ 
is given by a superposition $ \ket{\psi} = \sum_{M_{J}} C_{M_{J}} \ket{M_{J}}$.
Reading from the left, the first two columns account for $ \ket{\psi_{A}^{(0)}} $ and $ \ket{\psi_{A+1}^{(0)}} $.
These are called $A$-doublets of the CF spectra to underline their 
different structure compared to $ \ket{\psi_{B}^{(0)}} $ and $ \ket{\psi_{B+1}^{(0)}} $, the $B$-doublets listed in the fifth and sixth columns.
The ground states, $\ket{\psi_{0}^{(0)}}, \ket{\psi_{1}^{(0)}}$, belong to the $A$-type doublets 
(explicit values of the coefficients are given in the captions of each table).

In the third and fourth columns, the states $ \hat{\widetilde{V}}\ket{\psi_{A}^{(0)}} $ and $ \hat{\widetilde{V}}\ket{\psi_{A+1}^{(0)}} $,
obtained by applying the perturbation $ \hat{\widetilde{V}} $ to the $A$-doublets,
are given to facilitate the calculation of  $\widetilde{V}_{0,m}$ and $\widetilde{V}_{1,m}$, i.e., 
the coupling between the ground state doublet and the other CF states.
The perturbed states in the third and fourth column are expressed in terms of 
coefficients $j_{M}^{\pm}$ defined as:
\begin{equation}
	\begin{split}
		\hat{\widetilde{V}}	\ket{M_{J}}	&=	j_{M_{J}}^{+}	\ket{ M_{J} + 1 } + j_{M_{J}}^{-} \ket{ M_{J} - 1 }	,	\\[0.1cm]
						j_{M_{J}}^{\pm}	&=	e^{ \mp i\phi }	\sqrt{ J(J+1) - M_{J} (M_{J} \pm 1 ) }	.
	\end{split}
\label{eq:VtildeJpm}
\end{equation}
The $j_{M_{J}}^{\pm}$ only depend on the angle $\phi$ of the field in Eq.~\eqref{eq:Vtilde} and on the quantum numbers $J,M_{J}$.
Furthermore, from the general properties of the ladder operators, we have
\begin{equation}
	\begin{split}
		j_{M_{J}}^{\pm}	=		j_{-(M_{J}		\pm	1)	}^	{\pm}		\,	,
	\end{split}
	\label{eq:propJpm}
\end{equation}
which leads to characteristic symmetries of the $\widetilde{V}_{n,m}$ elements
that are key to determine the behaviour of the leading orders of the perturbative splitting of the ground state doublet. 
Since HTO is a non-Kramers system, Table~\ref{tab:HTOstates} also shows two kinds of singlets in the last two columns on the right.


\subsection{	HTO 	\label{sec:App_MatElemPer_HTOstates}	}

The crystal-field spectrum of HTO is made of 5 singlets and 6 doublets (see Fig.~\ref{fig:CFenergiesHTO}). 
As summarised in Table~\ref{tab:HTOstates}, there are two types of doublets, $ A $ and $ B $, and two types of singlets, $ s $ and $ s' $.

\subsubsection{	$A$-doublets		\label{sec:App_MatElemPer_AdoubletHTO}	}

The ground state doublet in HTO, $\ket{\psi_{0}^{(0)}} $ and $ \ket{\psi_{1}^{(0)}}$, is made up of $A$-type states, as defined in Table~\ref{tab:HTOstates}. 

It is then immediate to prove that $\widetilde{V}_{0,0}=0$ and $\widetilde{V}_{1,1}=0$ since in general 
$ \bra{\psi_{A}^{(0)}}\hat{\widetilde{V}}\ket{\psi_{A}^{(0)}}=0$ and $ \bra{\psi_{A+1}^{(0)}}\hat{\widetilde{V}}\ket{\psi_{A+1}^{(0)}}=0$. 
Namely, the first and the third column (and the second and fourth) of Table~\ref{tab:HTOstates} have trivially vanishing overlap. 

On the contrary the first and fourth (second and third) columns have a priori non-vanising overlap. In order to see that again  $\widetilde{V}_{1,0}=\widetilde{V}_{0,1}=0$ one ought to consider 
the explicit form of the matrix elements 
\begin{widetext}
\begin{equation}
	\bra{\psi_{A+1}^{(0)}}\hat{\widetilde{V}}\ket{\psi_{A}^{(0)}}
					=	a_{-7}a_{8}	\bigg(	j_{-7}^{-}-j_{8}^{-}	\bigg)
					+	a_{-4}a_{5}	\bigg(	j_{-4}^{-}-j_{5}^{-}	\bigg)
					+	a_{-1}a_{2}	\bigg(	j_{-1}^{-}-j_{2}^{-}	\bigg)
\, , 
	\label{eqA:V10HTO}
\end{equation}
\end{widetext} 
which vanishes because all elements within round brackets cancel out, according to Eq.~\eqref{eq:propJpm}. 

This shows not only that $\widetilde{V}_{0,0}=\widetilde{V}_{1,1}=\widetilde{V}_{1,0}=\widetilde{V}_{0,1}=0$,
accounting for the vanishing first order splitting in Eq.~\eqref{eq:Delta01Perturb} in the main text, but also 
that all matrix elements coupling the ground state with any other $A$-doublet of the CF spectrum have to be null. 
Summarising, Table~\ref{tab:HTOstates} and Eq.~\eqref{eqA:V10HTO} prove the general property 
$ \bra{\psi_{A'}^{(0)}}\hat{\widetilde{V}}\ket{\psi_{A}^{(0)}}=\bra{\psi_{A'+1}^{(0)}}\hat{\widetilde{V}}\ket{\psi_{A+1}^{(0)}}=\bra{\psi_{A'+1}^{(0)}}\hat{\widetilde{V}}\ket{\psi_{A}^{(0)}}=0 $
for any two doublets  $ A $ and $ A' $ in the CF spectrum of HTO.

\subsubsection{	$B$-doublets		\label{sec:App_MatElemPer_BdoubletHTO}	}

The other type of doublets in the CF spectrum of HTO are the $B$-doublets.
The matrix elements
$\bra{\psi_{B}^{(0)}}\hat{\widetilde{V}}\ket{\psi_{m}^{(0)}}$ 	
and 
$\bra{\psi_{B+1}^{(0)}}\hat{\widetilde{V}}\ket{\psi_{m}^{(0)}}$
are non zero for both states of the ground state doublet ($m=0,1$).
This is because in general the overlap between the perturbed $A$-states, in column three and four,
with the $B$-states, in columns five and six, is non-zero.
Here, for brevity, only the results for $\ket{\psi_{A}^{(0)}}$ are shown explicitly:
\begin{widetext} 
\begin{equation}
	\begin{split}
		\bra{\psi_{B}^{(0)}}\hat{\widetilde{V}}\ket{\psi_{A}^{(0)}}	&=	\sum_{M=2,5,8}			j_{M}^{-}	\bigg(  a_{M} b_{M-1}    + a_{-(M-1)} b_{-M}  \bigg)	
		 	 \\[0.1cm]
		\bra{\psi_{B+1}^{(0)}}\hat{\widetilde{V}}\ket{\psi_{A}^{(0)}}	&=	\sum_{M=2,5,8}	(-1)^{M}	j_{M}^{-}	\bigg(  a_{M}b_{-(M-1)}  - a_{-(M-1)}b_{M}     \bigg)	
		\, .
	\end{split}
	\label{eqA:HTO_VDD101}
\end{equation}
\end{widetext} 
Analougously one can show that also 
$ \bra{\psi_{B}^{(0)}}\hat{\widetilde{V}}\ket{\psi_{A+1}^{(0)}}$
and
$\bra{\psi_{B+1}^{(0)}}\hat{\widetilde{V}}\ket{\psi_{A+1}^{(0)}}$ are non zero.

Using their conjugation properties, one finds that 
\begin{equation}
	\begin{split}
		\bra{\psi_{B}^{(0)}}		\hat{\widetilde{V}}	\ket{\psi_{A+1}^{(0)}}		&=	 \bra{\psi_{A}^{(0)}}		\hat{\widetilde{V}}\ket{\psi_{B+1}^{(0)}} \\[0.1cm]
		\bra{\psi_{B+1}^{(0)}}		\hat{\widetilde{V}}	\ket{\psi_{A+1}^{(0)}}		&=	-\bra{\psi_{A}^{(0)}}		\hat{\widetilde{V}}\ket{\psi_{B}^{(0)}} 		.
	\end{split}
	\label{eqA:VDD:10:1HTO}
\end{equation}
whose implications, in the context of the ground state splitting, 
are discussed in Sec.~\ref{sec:MagnField_PerturbTheory} of the main text. 
Namely, the contribution to second order splitting due to $B$-doublets vanishes identically. 

Whereas for notational convenience we have worked with a given choice of eigenstates for both the GS doublet and excited state doublets, the main results are independent of it. For instance, one can verify with a few lines of algebra that the relations in Eq.~\eqref{eqA:VDD:10:1HTO} are invariant under generic basis transformations within each doublet involved.

\subsubsection{	Singlets 	\label{sec:App_MatElemPer_singletsHTO}	}

Another interesting feature of the CF eigenstates for HTO is the structure of the singlets $\ket{\psi_{s}^{(0)}}$ and $\ket{\psi_{s'}^{(0)}}$. 
These are shown respectively in the fifth and sixth column (from the left) of Table~\ref{tab:HTOstates}.
To avoid confusion, it is important to underline that, in general, $ s_{i} \neq s'_{j}$ for all $i,j$. 
The perturbative coupling of the singlets with the ground state doublet is non vanishing for both kind of singlets. 
Here, for brevity, we show explicitly only the matrix elements for 
$ \ket{\psi_{A}^{(0)}} $: 
\begin{widetext} 
\begin{equation}	\label{eqA:Vs0}
	\begin{split}
			\bra{\psi_{s}^{(0)}}\hat{\widetilde{V}}\ket{\psi_{A}^{(0)}}	&=	
			s_{3}\bigg(a_{2}j_{2}^{+}-a_{-4}j_{-4}^{+}\bigg)+s_{6}\bigg(a_{5}j_{5}^{+} + a_{-7}j_{-7}^{+}\bigg)+s_{0}a_{-1}j_{-1}^{+}	
			\,	,	\\[0.1cm]
			\bra{\psi_{s'}^{(0)}}\hat{\widetilde{V}}\ket{\psi_{A}^{(0)}}	&=	
			s'_{3}\bigg(a_{2}j_{2}^{+}+a_{-4}j_{-4}^{+}\bigg)+s'_{6}\bigg(a_{5}j_{5}^{+}-a_{-7}j_{-7}^{+}\bigg)	
			\,	.
	\end{split}
\end{equation}
\end{widetext} 
Analogously, it is straightforward to show that 
$\bra{\psi_{s}^{(0)}}\hat{\widetilde{V}}\ket{\psi_{A+1}^{(0)}} \neq 0$ and  $\bra{\psi_{s'}^{(0)}}\hat{\widetilde{V}}\ket{\psi_{A+1}^{(0)}} \neq 0$. 

The matrix elements coupling the ground state doublet to the singlets 
provide the only non-vanishing second-order contribution to the ground state splitting in Eq.~\eqref{eq:Delta01Perturb}, 
marking the difference in the power law dependence found for HTO and DTO, as discussed in Sec.~\ref{sec:MagnField_PerturbTheory}.

\subsection{	DTO 	\label{sec:App_MatElemPer_DTOstates}	}

All the energy levels in the crystal-field spectrum of DTO are doublets (see Fig.~\ref{fig:CFenergiesDTO}).
In Table~\ref{tab:DTOstates} these are distinguished into $ A $ and $ B $ doublets, in analogy with HTO.

\subsubsection{	$A$-doublets		\label{sec:App_MatElemPer_AdoubletDTO}	}

The two basis states, $\ket{\psi_{0}^{(0)}} $ and $ \ket{\psi_{1}^{(0)}}$, of the ground doublet of the DTO crystal-field spectrum are of type $A$, as defined in Table~\ref{tab:DTOstates}. 
It is then straightforward to verify that  $\widetilde{V}_{0,0}=\widetilde{V}_{1,1}=\widetilde{V}_{0,1}=\widetilde{V}_{0,1}=0$ since in general 
%
%
\begin{equation}
	\begin{split}
		\bra{\psi_{A'}^{(0)}}	\hat{\widetilde{V}}\ket{\psi_{A}^{(0)}}		=	
		\bra{\psi_{A'+1}^{(0)}}\hat{\widetilde{V}}\ket{\psi_{A+1}^{(0)}}	=
		\bra{\psi_{A'+1}^{(0)}}\hat{\widetilde{V}}\ket{\psi_{A}^{(0)}}		=	0,
	\end{split}
	\label{eqA:V10DTO}
\end{equation}
%
by comparing the first pair and the second pair of columns in Table~\ref{tab:DTOstates}. 
The null matrix elements in Eq.~\eqref{eqA:V10DTO} are responsible for the vanishing 
of the first order contribution to the ground state splitting in Eq.~\eqref{eq:Delta01Perturb}.

\subsubsection{	$B$-doublets		\label{sec:App_MatElemPer_BdoubletDTO}	}

As for HTO, also for DTO the matrix elements coupling the $ A $ and $ B $ doublets are non-vanishing: 
\begin{widetext} 
\begin{equation}
	\begin{split}
		\bra{\psi_{B}^{(0)}}\hat{\widetilde{V}}\ket{\psi_{A}^{(0)}}	
		&=	\sum_{M=-9/2}^{15/2}			
		\bigg(  
			j_{M}^{-} \, a_{M} b_{M-1}  +  j_{-M}^{+} \, a_{-M} b_{-(M-1)}  
		\bigg)	
			 \\[0.1cm]
		\bra{\psi_{B+1}^{(0)}}\hat{\widetilde{V}}\ket{\psi_{A}^{(0)}}	
		&=	\sum_{M=-9/2}^{15/2} (-1)^{M+\frac{1}{2}}	
		\bigg(  
			j_{M}^{-} \, a_{M} b_{-(M-1)}  +  j_{-M}^{+} \, a_{-M} b_{M-1}  
		\bigg)	
		\, ,
	\end{split}
	\label{eqA:DTO_VDD101}
\end{equation}
\end{widetext} 
where the sum over the $ M $ quantum numbers runs, from $ -9/2 $  to $ 15/2 $, in intervals of 3 ($ M=-9/2,\,-3/2,\,3/2,\,9/2,\,15/2 $).

Their conjugation properties give: 
\begin{equation}
	\begin{split}
		\bra{\psi_{B}^{(0)}}		\hat{\widetilde{V}}	\ket{\psi_{A+1}^{(0)}}		&=	-\bra{\psi_{A}^{(0)}}		\hat{\widetilde{V}}\ket{\psi_{B+1}^{(0)}} \\[0.1cm]
		\bra{\psi_{B+1}^{(0)}}		\hat{\widetilde{V}}	\ket{\psi_{A+1}^{(0)}}		&=	 \bra{\psi_{A}^{(0)}}		\hat{\widetilde{V}}\ket{\psi_{B}^{(0)}} 		,
	\end{split}
	\label{eqA:VDD:10:1DTO}
\end{equation}
whose signs are opposite to the case of HTO in Eq.~\eqref{eqA:VDD:10:1HTO}. 
Similarly to the case of HTO however, Eqs.~\eqref{eqA:VDD:10:1DTO} give a vanishing second order contribution to the splitting of the ground state doublet. Since there are no singlets in DTO, no splitting at all takes place to second order. 

Because all the matrix elements in Eq.~\eqref{eqA:V10DTO} are null,
the matrix elements in Eq.~\eqref{eqA:VDD:10:1DTO} are the only ones ultimately responsible for the DTO energy splitting, which takes place to third order in (transverse field) perturbation theory, as illustrated in Sec.~\ref{sec:MagnField}. 

We stress once again that, whereas for notational convenience we have worked with a given choice of eigenstates for both the GS doublet and excited state doublets, the main results are independent of it. For instance, one can readily verify that the relations in Eq.~\eqref{eqA:VDD:10:1DTO} are invariant under generic basis transformations within each doublet involved. 



\begin{thebibliography}{99}

\expandafter\ifx\csname natexlab\endcsname\relax\def\natexlab#1{#1}\fi
\expandafter\ifx\csname bibnamefont\endcsname\relax
  \def\bibnamefont#1{#1}\fi
\expandafter\ifx\csname bibfnamefont\endcsname\relax
  \def\bibfnamefont#1{#1}\fi
\expandafter\ifx\csname citenamefont\endcsname\relax
  \def\citenamefont#1{#1}\fi
\expandafter\ifx\csname url\endcsname\relax
  \def\url#1{\texttt{#1}}\fi
\expandafter\ifx\csname urlprefix\endcsname\relax\def\urlprefix{URL }\fi
\providecommand{\bibinfo}[2]{#2}
\providecommand{\eprint}[2][]{\url{#2}}

\bibitem[{\citenamefont{Bramwell and Gingras}(2001)}]{Bramwell:2001}
\bibinfo{author}{\bibfnamefont{S.~T.} \bibnamefont{Bramwell}} \bibnamefont{and}
  \bibinfo{author}{\bibfnamefont{M.~J.~P.} \bibnamefont{Gingras}},
  \bibinfo{journal}{Science} \textbf{\bibinfo{volume}{294}},
  \bibinfo{pages}{1495} (\bibinfo{year}{2001}).

\bibitem[{\citenamefont{Rau and Gingras}(2015)}]{Rau:2015}
\bibinfo{author}{\bibfnamefont{J.~G.} \bibnamefont{Rau}} \bibnamefont{and}
  \bibinfo{author}{\bibfnamefont{M.~J.~P.} \bibnamefont{Gingras}},
  \bibinfo{journal}{arXiv} \textbf{\bibinfo{volume}{1503.04808}}
  (\bibinfo{year}{2015}).

\bibitem[{\citenamefont{Castelnovo et~al.}(2008)\citenamefont{Castelnovo,
  Moessner, and Sondhi}}]{Castelnovo:2008}
\bibinfo{author}{\bibfnamefont{C.}~\bibnamefont{Castelnovo}},
  \bibinfo{author}{\bibfnamefont{R.}~\bibnamefont{Moessner}}, \bibnamefont{and}
  \bibinfo{author}{\bibfnamefont{S.~L.} \bibnamefont{Sondhi}},
  \bibinfo{journal}{Nature} \textbf{\bibinfo{volume}{451}}, \bibinfo{pages}{42}
  (\bibinfo{year}{2008}).

\bibitem[{\citenamefont{Castelnovo et~al.}(2012)\citenamefont{Castelnovo,
  Moessner, and Sondhi}}]{Castelnovo:2012}
\bibinfo{author}{\bibfnamefont{C.}~\bibnamefont{Castelnovo}},
  \bibinfo{author}{\bibfnamefont{R.}~\bibnamefont{Moessner}}, \bibnamefont{and}
  \bibinfo{author}{\bibfnamefont{S.}~\bibnamefont{Sondhi}},
  \bibinfo{journal}{Annual Review of Condensed Matter Physics}
  \textbf{\bibinfo{volume}{3}}, \bibinfo{pages}{35} (\bibinfo{year}{2012}).

\bibitem[{\citenamefont{Snyder et~al.}(2001)\citenamefont{Snyder, Slusky, Cava,
  and Schiffer}}]{Snyder:2001}
\bibinfo{author}{\bibfnamefont{J.}~\bibnamefont{Snyder}},
  \bibinfo{author}{\bibfnamefont{J.~S.} \bibnamefont{Slusky}},
  \bibinfo{author}{\bibfnamefont{R.~J.} \bibnamefont{Cava}}, \bibnamefont{and}
  \bibinfo{author}{\bibfnamefont{P.}~\bibnamefont{Schiffer}},
  \bibinfo{journal}{Nature} \textbf{\bibinfo{volume}{413}}, \bibinfo{pages}{48}
  (\bibinfo{year}{2001}).

\bibitem[{\citenamefont{Ehlers et~al.}(2003)\citenamefont{Ehlers, Cornelius,
  Orend{\'a}c, Kajnakov{\'a}, Fennell, Bramwell, and Gardner}}]{Ehlers:2003}
\bibinfo{author}{\bibfnamefont{G.}~\bibnamefont{Ehlers}},
  \bibinfo{author}{\bibfnamefont{A.~L.} \bibnamefont{Cornelius}},
  \bibinfo{author}{\bibfnamefont{M.}~\bibnamefont{Orend{\'a}c}},
  \bibinfo{author}{\bibfnamefont{M.}~\bibnamefont{Kajnakov{\'a}}},
  \bibinfo{author}{\bibfnamefont{T.}~\bibnamefont{Fennell}},
  \bibinfo{author}{\bibfnamefont{S.~T.} \bibnamefont{Bramwell}},
  \bibnamefont{and} \bibinfo{author}{\bibfnamefont{J.~S.}
  \bibnamefont{Gardner}}, \bibinfo{journal}{Journal of Physics: Condensed
  Matter} \textbf{\bibinfo{volume}{15}}, \bibinfo{pages}{L9}
  (\bibinfo{year}{2003}).

\bibitem[{\citenamefont{Snyder et~al.}(2004)\citenamefont{Snyder, Ueland,
  Slusky, Karunadasa, Cava, and Schiffer}}]{Snyder:2004}
\bibinfo{author}{\bibfnamefont{J.}~\bibnamefont{Snyder}},
  \bibinfo{author}{\bibfnamefont{B.~G.} \bibnamefont{Ueland}},
  \bibinfo{author}{\bibfnamefont{J.~S.} \bibnamefont{Slusky}},
  \bibinfo{author}{\bibfnamefont{H.}~\bibnamefont{Karunadasa}},
  \bibinfo{author}{\bibfnamefont{R.~J.} \bibnamefont{Cava}}, \bibnamefont{and}
  \bibinfo{author}{\bibfnamefont{P.}~\bibnamefont{Schiffer}},
  \bibinfo{journal}{Physical Review B} \textbf{\bibinfo{volume}{69}},
  \bibinfo{pages}{064414} (\bibinfo{year}{2004}).

\bibitem[{\citenamefont{Matsuhira et~al.}(2011)\citenamefont{Matsuhira,
  Paulsen, Lhotel, Sekine, Hiroi, and Takagi}}]{Matsuhira:2011}
\bibinfo{author}{\bibfnamefont{K.}~\bibnamefont{Matsuhira}},
  \bibinfo{author}{\bibfnamefont{C.}~\bibnamefont{Paulsen}},
  \bibinfo{author}{\bibfnamefont{E.}~\bibnamefont{Lhotel}},
  \bibinfo{author}{\bibfnamefont{C.}~\bibnamefont{Sekine}},
  \bibinfo{author}{\bibfnamefont{Z.}~\bibnamefont{Hiroi}}, \bibnamefont{and}
  \bibinfo{author}{\bibfnamefont{S.}~\bibnamefont{Takagi}},
  \bibinfo{journal}{Journal of the Physical Society of Japan}
  \textbf{\bibinfo{volume}{80}}, \bibinfo{pages}{123711}
  (\bibinfo{year}{2011}).

\bibitem[{\citenamefont{Jaubert and Holdsworth}(2009)}]{Jaubert:2009}
\bibinfo{author}{\bibfnamefont{L.~D.~C.} \bibnamefont{Jaubert}}
  \bibnamefont{and} \bibinfo{author}{\bibfnamefont{P.~C.~W.}
  \bibnamefont{Holdsworth}}, \bibinfo{journal}{Nat Phys}
  \textbf{\bibinfo{volume}{5}}, \bibinfo{pages}{258} (\bibinfo{year}{2009}).

\bibitem[{\citenamefont{Ryzhkin}(2005)}]{Ryzhkin:2005}
\bibinfo{author}{\bibfnamefont{I.}~\bibnamefont{Ryzhkin}},
  \bibinfo{journal}{Journal of Experimental and Theoretical Physics}
  \textbf{\bibinfo{volume}{101}}, \bibinfo{pages}{481} (\bibinfo{year}{2005}),
  ISSN \bibinfo{issn}{1063-7761}.

\bibitem[{\citenamefont{Castelnovo et~al.}(2010)\citenamefont{Castelnovo,
  Moessner, and Sondhi}}]{Castelnovo:2010}
\bibinfo{author}{\bibfnamefont{C.}~\bibnamefont{Castelnovo}},
  \bibinfo{author}{\bibfnamefont{R.}~\bibnamefont{Moessner}}, \bibnamefont{and}
  \bibinfo{author}{\bibfnamefont{S.~L.} \bibnamefont{Sondhi}},
  \bibinfo{journal}{Physical Review Letters} \textbf{\bibinfo{volume}{104}},
  \bibinfo{pages}{107201} (\bibinfo{year}{2010}).

\bibitem[{\citenamefont{Mostame et~al.}(2014)\citenamefont{Mostame, Castelnovo,
  Moessner, and Sondhi}}]{Mostame:2014}
\bibinfo{author}{\bibfnamefont{S.}~\bibnamefont{Mostame}},
  \bibinfo{author}{\bibfnamefont{C.}~\bibnamefont{Castelnovo}},
  \bibinfo{author}{\bibfnamefont{R.}~\bibnamefont{Moessner}}, \bibnamefont{and}
  \bibinfo{author}{\bibfnamefont{S.~L.} \bibnamefont{Sondhi}},
  \bibinfo{journal}{Proceedings of the National Academy of Sciences}
  \textbf{\bibinfo{volume}{111}}, \bibinfo{pages}{640} (\bibinfo{year}{2014}).

\bibitem[{\citenamefont{Isakov et~al.}(2005)\citenamefont{Isakov, Moessner, and
  Sondhi}}]{Isakov:2005}
\bibinfo{author}{\bibfnamefont{S.~V.} \bibnamefont{Isakov}},
  \bibinfo{author}{\bibfnamefont{R.}~\bibnamefont{Moessner}}, \bibnamefont{and}
  \bibinfo{author}{\bibfnamefont{S.~L.} \bibnamefont{Sondhi}},
  \bibinfo{journal}{Physical Review Letters} \textbf{\bibinfo{volume}{95}},
  \bibinfo{pages}{217201} (\bibinfo{year}{2005}).

\bibitem[{\citenamefont{Ramirez et~al.}(1999)\citenamefont{Ramirez, Hayashi,
  Cava, Siddharthan, and Shastry}}]{Ramirez:1999}
\bibinfo{author}{\bibfnamefont{A.~P.} \bibnamefont{Ramirez}},
  \bibinfo{author}{\bibfnamefont{A.}~\bibnamefont{Hayashi}},
  \bibinfo{author}{\bibfnamefont{R.~J.} \bibnamefont{Cava}},
  \bibinfo{author}{\bibfnamefont{R.~B.} \bibnamefont{Siddharthan}},
  \bibnamefont{and} \bibinfo{author}{\bibfnamefont{S.}~\bibnamefont{Shastry}},
  \bibinfo{journal}{Nature} \textbf{\bibinfo{volume}{399}},
  \bibinfo{pages}{333} (\bibinfo{year}{1999}).

\bibitem[{\citenamefont{Paulsen et~al.}(2014)\citenamefont{Paulsen, Jackson,
  Lhotel, Canals, Prabhakaran, Matsuhira, Giblin, and Bramwell}}]{Paulsen:2014}
\bibinfo{author}{\bibfnamefont{C.}~\bibnamefont{Paulsen}},
  \bibinfo{author}{\bibfnamefont{M.~J.} \bibnamefont{Jackson}},
  \bibinfo{author}{\bibfnamefont{E.}~\bibnamefont{Lhotel}},
  \bibinfo{author}{\bibfnamefont{B.}~\bibnamefont{Canals}},
  \bibinfo{author}{\bibfnamefont{D.}~\bibnamefont{Prabhakaran}},
  \bibinfo{author}{\bibfnamefont{K.}~\bibnamefont{Matsuhira}},
  \bibinfo{author}{\bibfnamefont{S.~R.} \bibnamefont{Giblin}},
  \bibnamefont{and} \bibinfo{author}{\bibfnamefont{S.~T.}
  \bibnamefont{Bramwell}}, \bibinfo{journal}{Nat Phys}
  \textbf{\bibinfo{volume}{10}}, \bibinfo{pages}{135} (\bibinfo{year}{2014}).

\bibitem[{\citenamefont{Pan et~al.}(2015)\citenamefont{Pan, Laurita, Ross,
  Kermarrec, Gaulin, and P.}}]{Pan:arX2015}
\bibinfo{author}{\bibfnamefont{L.}~\bibnamefont{Pan}},
  \bibinfo{author}{\bibfnamefont{N.~J.} \bibnamefont{Laurita}},
  \bibinfo{author}{\bibfnamefont{K.~A.} \bibnamefont{Ross}},
  \bibinfo{author}{\bibfnamefont{E.}~\bibnamefont{Kermarrec}},
  \bibinfo{author}{\bibfnamefont{B.~D.} \bibnamefont{Gaulin}},
  \bibnamefont{and} \bibinfo{author}{\bibfnamefont{A.~N.} \bibnamefont{P.}},
  \bibinfo{journal}{arXiv} \textbf{\bibinfo{volume}{1501.05638}}
  (\bibinfo{year}{2015}).

\bibitem[{\citenamefont{Tokiwa et~al.}(2015)\citenamefont{Tokiwa, Yamashita,
  Udagawa, Kittaka, Sakakibara, Terazawa, Shimoyama, Terashima, Yasui,
  Shibauchi et~al.}}]{Tokiwa:arX2015}
\bibinfo{author}{\bibfnamefont{Y.}~\bibnamefont{Tokiwa}},
  \bibinfo{author}{\bibfnamefont{T.}~\bibnamefont{Yamashita}},
  \bibinfo{author}{\bibfnamefont{M.}~\bibnamefont{Udagawa}},
  \bibinfo{author}{\bibfnamefont{S.}~\bibnamefont{Kittaka}},
  \bibinfo{author}{\bibfnamefont{T.}~\bibnamefont{Sakakibara}},
  \bibinfo{author}{\bibfnamefont{D.}~\bibnamefont{Terazawa}},
  \bibinfo{author}{\bibfnamefont{Y.}~\bibnamefont{Shimoyama}},
  \bibinfo{author}{\bibfnamefont{Y.~T.} \bibnamefont{Terashima}},
  \bibinfo{author}{\bibfnamefont{Y.}~\bibnamefont{Yasui}},
  \bibinfo{author}{\bibfnamefont{T.}~\bibnamefont{Shibauchi}},
  \bibnamefont{et~al.}, \bibinfo{journal}{arXiv}
  \textbf{\bibinfo{volume}{1504.02199}} (\bibinfo{year}{2015}).

\bibitem[{\citenamefont{Petrova et~al.}(2015)\citenamefont{Petrova, Moessner,
  and Sondhi}}]{Petrova:arX2015}
\bibinfo{author}{\bibfnamefont{O.}~\bibnamefont{Petrova}},
  \bibinfo{author}{\bibfnamefont{R.}~\bibnamefont{Moessner}}, \bibnamefont{and}
  \bibinfo{author}{\bibfnamefont{S.~L.} \bibnamefont{Sondhi}},
  \bibinfo{journal}{arXiv} \textbf{\bibinfo{volume}{1501.02445}}
  (\bibinfo{year}{2015}).

\bibitem[{\citenamefont{Sala et~al.}(2012)\citenamefont{Sala, Castelnovo,
  Moessner, Sondhi, Kitagawa, Takigawa, Higashinaka, and Maeno}}]{Sala:2012}
\bibinfo{author}{\bibfnamefont{G.}~\bibnamefont{Sala}},
  \bibinfo{author}{\bibfnamefont{C.}~\bibnamefont{Castelnovo}},
  \bibinfo{author}{\bibfnamefont{R.}~\bibnamefont{Moessner}},
  \bibinfo{author}{\bibfnamefont{S.~L.} \bibnamefont{Sondhi}},
  \bibinfo{author}{\bibfnamefont{K.}~\bibnamefont{Kitagawa}},
  \bibinfo{author}{\bibfnamefont{M.}~\bibnamefont{Takigawa}},
  \bibinfo{author}{\bibfnamefont{R.}~\bibnamefont{Higashinaka}},
  \bibnamefont{and} \bibinfo{author}{\bibfnamefont{Y.}~\bibnamefont{Maeno}},
  \bibinfo{journal}{Physical Review Letters} \textbf{\bibinfo{volume}{108}},
  \bibinfo{pages}{217203} (\bibinfo{year}{2012}).

\bibitem[{\citenamefont{Subramanian et~al.}(1993)\citenamefont{Subramanian,
  Sleight, Karl A.~Gschneidner, and Eyring}}]{Subramanian:1993}
\bibinfo{author}{\bibfnamefont{M.~A.} \bibnamefont{Subramanian}},
  \bibinfo{author}{\bibfnamefont{A.~W.} \bibnamefont{Sleight}},
  \bibinfo{author}{\bibfnamefont{J.}~\bibnamefont{Karl A.~Gschneidner}},
  \bibnamefont{and} \bibinfo{author}{\bibfnamefont{L.}~\bibnamefont{Eyring}},
  \emph{\bibinfo{title}{Chapter 107 Rare earth pyrochlores}}
  (\bibinfo{publisher}{Elsevier}, \bibinfo{year}{1993}),
  vol.~\bibinfo{volume}{16}, pp. \bibinfo{pages}{225--248}, ISBN
  \bibinfo{isbn}{0168-1273}.

\bibitem[{\citenamefont{Gardner et~al.}(2010)\citenamefont{Gardner, Gingras,
  and Greedan}}]{Gardner:2010}
\bibinfo{author}{\bibfnamefont{J.~S.} \bibnamefont{Gardner}},
  \bibinfo{author}{\bibfnamefont{M.~J.~P.} \bibnamefont{Gingras}},
  \bibnamefont{and} \bibinfo{author}{\bibfnamefont{J.~E.}
  \bibnamefont{Greedan}}, \bibinfo{journal}{Reviews of Modern Physics}
  \textbf{\bibinfo{volume}{82}}, \bibinfo{pages}{53} (\bibinfo{year}{2010}).

\bibitem[{\citenamefont{Abragam and Bleaney}(1987)}]{AbragamBleaneyBook:1987}
\bibinfo{author}{\bibfnamefont{A.}~\bibnamefont{Abragam}} \bibnamefont{and}
  \bibinfo{author}{\bibfnamefont{B.}~\bibnamefont{Bleaney}},
  \emph{\bibinfo{title}{Electron Paramagnetic Resonance of Transition Ions}}
  (\bibinfo{publisher}{Dover Publications Inc.}, \bibinfo{year}{1987}).

\bibitem[{\citenamefont{H{\"u}fner}(1978)}]{Hufner:1978}
\bibinfo{author}{\bibfnamefont{S.}~\bibnamefont{H{\"u}fner}},
  \emph{\bibinfo{title}{Optical spectra of transparent rare earth compounds}}
  (\bibinfo{publisher}{Academic Press}, \bibinfo{year}{1978}), ISBN
  \bibinfo{isbn}{9780123604507}.

\bibitem[{\citenamefont{Takegahara}(2000)}]{Takegahara:2000}
\bibinfo{author}{\bibfnamefont{K.}~\bibnamefont{Takegahara}},
  \bibinfo{journal}{Journal of the Physical Society of Japan}
  \textbf{\bibinfo{volume}{69}}, \bibinfo{pages}{1572} (\bibinfo{year}{2000}).

\bibitem[{\citenamefont{Hutchings et~al.}(1964)\citenamefont{Hutchings, Seitz,
  and Turnbull}}]{Hutchings:1964}
\bibinfo{author}{\bibfnamefont{M.}~\bibnamefont{Hutchings}},
  \bibinfo{author}{\bibfnamefont{F.}~\bibnamefont{Seitz}}, \bibnamefont{and}
  \bibinfo{author}{\bibfnamefont{D.}~\bibnamefont{Turnbull}},
  \textbf{\bibinfo{volume}{16}}, \bibinfo{pages}{227} (\bibinfo{year}{1964}).

\bibitem[{\citenamefont{Rosenkranz et~al.}(2000)\citenamefont{Rosenkranz,
  Ramirez, Hayashi, Cava, Siddharthan, and Shastry}}]{Rosenkranz:2000}
\bibinfo{author}{\bibfnamefont{S.}~\bibnamefont{Rosenkranz}},
  \bibinfo{author}{\bibfnamefont{A.~P.} \bibnamefont{Ramirez}},
  \bibinfo{author}{\bibfnamefont{A.}~\bibnamefont{Hayashi}},
  \bibinfo{author}{\bibfnamefont{R.~J.} \bibnamefont{Cava}},
  \bibinfo{author}{\bibfnamefont{R.}~\bibnamefont{Siddharthan}},
  \bibnamefont{and} \bibinfo{author}{\bibfnamefont{B.~S.}
  \bibnamefont{Shastry}}, \bibinfo{journal}{Journal of Applied Physics}
  \textbf{\bibinfo{volume}{87}}, \bibinfo{pages}{5914} (\bibinfo{year}{2000}).

\bibitem[{\citenamefont{Malkin et~al.}(2010)\citenamefont{Malkin, Lummen, van
  Loosdrecht, Dhalenne, and Zakirov}}]{Malkin:2010}
\bibinfo{author}{\bibfnamefont{B.}~\bibnamefont{Malkin}},
  \bibinfo{author}{\bibfnamefont{T.}~\bibnamefont{Lummen}},
  \bibinfo{author}{\bibfnamefont{P.}~\bibnamefont{van Loosdrecht}},
  \bibinfo{author}{\bibfnamefont{G.}~\bibnamefont{Dhalenne}}, \bibnamefont{and}
  \bibinfo{author}{\bibfnamefont{A.}~\bibnamefont{Zakirov}},
  \bibinfo{journal}{Journal of physics. Condensed matter : an Institute of
  Physics journal} \textbf{\bibinfo{volume}{22}}, \bibinfo{pages}{276003}
  (\bibinfo{year}{2010}).

\bibitem[{\citenamefont{Erfanifam et~al.}(2014)\citenamefont{Erfanifam,
  Zherlitsyn, Yasin, Skourski, Wosnitza, Zvyagin, McClarty, Moessner,
  Balakrishnan, and Petrenko}}]{Erfanifam:2014}
\bibinfo{author}{\bibfnamefont{S.}~\bibnamefont{Erfanifam}},
  \bibinfo{author}{\bibfnamefont{S.}~\bibnamefont{Zherlitsyn}},
  \bibinfo{author}{\bibfnamefont{S.}~\bibnamefont{Yasin}},
  \bibinfo{author}{\bibfnamefont{Y.}~\bibnamefont{Skourski}},
  \bibinfo{author}{\bibfnamefont{J.}~\bibnamefont{Wosnitza}},
  \bibinfo{author}{\bibfnamefont{A.~A.} \bibnamefont{Zvyagin}},
  \bibinfo{author}{\bibfnamefont{P.}~\bibnamefont{McClarty}},
  \bibinfo{author}{\bibfnamefont{R.}~\bibnamefont{Moessner}},
  \bibinfo{author}{\bibfnamefont{G.}~\bibnamefont{Balakrishnan}},
  \bibnamefont{and} \bibinfo{author}{\bibfnamefont{O.~A.}
  \bibnamefont{Petrenko}}, \bibinfo{journal}{Physical Review B}
  \textbf{\bibinfo{volume}{90}}, \bibinfo{pages}{064409}
  (\bibinfo{year}{2014}).

\bibitem[{\citenamefont{Stevens}(1952)}]{Stevens:1952}
\bibinfo{author}{\bibfnamefont{K.~W.~H.} \bibnamefont{Stevens}},
  \bibinfo{journal}{Proceedings of the Physical Society. Section A}
  \textbf{\bibinfo{volume}{65}}, \bibinfo{pages}{209} (\bibinfo{year}{1952}).

\bibitem[{\citenamefont{Mulak and Gajek}(2000)}]{Mulak:2000}
\bibinfo{author}{\bibfnamefont{J.}~\bibnamefont{Mulak}} \bibnamefont{and}
  \bibinfo{author}{\bibfnamefont{Z.}~\bibnamefont{Gajek}},
  \emph{\bibinfo{title}{The Effective Crystal Field Potential}}
  (\bibinfo{publisher}{Elsevier Science}, \bibinfo{year}{2000}), ISBN
  \bibinfo{isbn}{9780080530710}.

\bibitem[{\citenamefont{Bransden and Joachain}(2000)}]{Bransden:2000}
\bibinfo{author}{\bibfnamefont{B.~H.} \bibnamefont{Bransden}} \bibnamefont{and}
  \bibinfo{author}{\bibfnamefont{C.~J.} \bibnamefont{Joachain}},
  \emph{\bibinfo{title}{Quantum Mechanics, 2nd edition}}
  (\bibinfo{publisher}{Addison-Wesley}, \bibinfo{year}{2000}).

\bibitem[{\citenamefont{Tomasello}(2014)}]{TomaselloPhDThesis}
\bibinfo{author}{\bibfnamefont{B.}~\bibnamefont{Tomasello}}, Ph.D. thesis,
  \bibinfo{school}{School of Physical Sciences, University of Kent}
  (\bibinfo{year}{2014}).




\end{thebibliography}

\end{document}